\documentclass[epsf,prb,twocolumn,showpacs]{revtex4-1}
\usepackage[pdftex]{graphicx}
\usepackage{dcolumn}
\usepackage{bm}
\usepackage{epsfig}
\usepackage{latexsym}
\usepackage{amsmath}
\usepackage{amsfonts}
\usepackage{amssymb}
\usepackage{color}
\usepackage{array}

\usepackage{setspace}

\newcommand{\<}{\langle}
\newcommand{\e}{\varepsilon}
\newcommand{\up}{\uparrow}
\newcommand{\down}{\downarrow}

\renewcommand{\>}{\rangle}
\renewcommand{\(}{\left(}
\renewcommand{\)}{\right)}
\renewcommand{\[}{\left[}
\renewcommand{\]}{\right]}
\renewcommand{\v}[1]{\mathbf{#1}} % \v -> vector (bf)

\newcommand{\bs}[1]{\boldsymbol{#1}}

\begin{document}
\pdfoutput=1

%%%% Title %%%%
\title{Mechanisms for Sub-Gap Optical Conductivity in Herbertsmithite}
\author{Andrew C. Potter}\author{T. Senthil}\author{Patrick A. Lee}
\affiliation{Massachusetts Institute of Technology 77 Massachusetts Ave. Cambridge, MA 02139}
\date{Nov 1, 2012 (Started)}
\begin{abstract}
Recent terahertz conductivity measurements observed low-power-law frequency dependence of optical conduction within the Mott gap of the Kagome lattice spin-liquid candidate Herbertsmithite. We investigate mechanisms for this observed sub-gap conductivity for two possible scenarios in which the ground-state is described by: 1) a U(1) Dirac spin-liquid with emergent fermionic spinons or 2) a nearly critical $Z_2$ spin-liquid in the vicinity of a continuous quantum phase transition to magnetic order.  We identify new mechanisms for optical-absorption via magneto-elastic effects and spin-orbit coupling.  In addition, for the Dirac spin-liquid scenario, we establish an explicit microscopic origin for previously proposed absorption mechanisms based on slave-particle effective field theory descriptions. 
\end{abstract}
\maketitle

\section{Introduction}
The spin-1/2 antiferromagnetic Heisenberg model on the Kagome lattice is thought to have a spin-liquid ground-state due to the interplay of geometric frustration and large quantum fluctuations of spin-1/2 moments\cite{Elser1,Marston,SachdevSPN,Elser2,Lecheminant,YanHuseWhite}.  This model is (approximately) realized in the material Herbertsmithite (ZnCu$_3$(OH)$_6$Cl$_2$), which consists of layers of Cu$^{2+}$ ions with  antiferromagnetically interacting spin-1/2 magnetic moments arranged in a structurally perfect Kagome lattice\cite{Shores}.  There is compelling evidence that Herbertsmithite has a spin-liquid ground-state. It shows no signs of magnetic order down to the lowest accessible temperatures $\approx 30$mK, well below the magnetic exchange scale $J\approx 190$K. Instead, at low-temperatures, it exhibits power law temperature dependence of spin-susceptibility $\chi_s$ and heat capacity $C_v$\cite{Helton}, as well as a Knight shift that tends to a constant at low T.\cite{Imai,Olariu}  These features are suggestive of gapless (or at least nearly gapless) spinon excitations.  Moreover, recent neutron measurements show a broad spectrum of spin-excitations consistent with a low-energy spinon continuum\cite{NeutronContinuum}.  

While Herbertsmithite is a promising spin-liquid candidate, the detailed nature of its putative spin-liquid ground-state remains mysterious.  Evidence for gapless spinon excitations led to early suggestions that the material might realize a U(1) Dirac spin-liquid (DSL) with fermionic spinons whose low-energy dispersion consists of two Dirac cones\cite{RanVMC,HermeleProperties}.  Projected wave-function studies showed that spin-wave-functions obtained from the DSL mean-field ansatz have very good energy with no variational parameters\cite{RanVMC} (even lower energies can be achieved by applying a few Lanczos steps to the DSL wave-function\cite{Becca}).  However, recent advances in density-matrix renormalization group (DMRG) techniques have enabled simulation of wide 1D strips, which suggest that the ground state of the Kagome Heisenberg model is instead a $Z_2$ spin-liquid with a sizeable spin-gap\cite{YanHuseWhite}.  The issue is complicated by the observation that the $Z_2$ spin-liquid found in DMRG for is very sensitive to small perturbations away from the pure Heisenberg model\cite{WhiteJ1J2}.  In light of this observation, it is quite plausible that additional ingredients such as disorder or spin-orbit coupling play an important role in determining the ground-state properties of Herbertsmithite.

These theoretical difficulties highlight the need for new experimental probes that can more directly measure the properties of emergent spinon excitations.  For example, it was recently proposed that spin-orbit coupling could enable neutron scattering measurements to detect the gapless emergent gauge fluctuations present in the U(1) Dirac spin-liquid theory\cite{LeeNagaosaNeutron}.

AC electrical conductivity measurements offer another route for measuring the structure of low-energy spin excitations.  Despite the host material being a strong Mott insulator, it was previously pointed out for a U(1) spin-liquid that virtual charge-fluctuations can enable power-law absorption deep within the Mott gap\cite{IoffeLarkin,NgLee}.  Indeed, in-gap optical absorption has been observed in organic spin-liquid candidates\cite{Elsasser}, but the data does not go to very low frequencies.  These suggestions have led to recent measurements of the electromagnetic properties of single-crystal samples of Herbertsmithite, at terahertz frequencies\cite{Pilon}.  This work finds a low-power-law frequency dependence to AC conductivity inside the Mott gap, whose magnitude increases as temperature is lowered.  

Inspired by this work, we revisit the issue of sub-gap AC electrical conductivity of candidate spin-liquid states for Herbertsmithite. We consider two scenarios in which Herbertsmithte forms either 1) a U(1) Dirac spin-liquid with fermionic spinons and an emergent U(1) gauge field or 2) a nearly gapless $Z_2$ spin-liquid in the vicinity of a quantum phase transition to non-collinear antiferromagnetic (AFM) order.  Within these scenarios, we find three distinct mechanisms for power-law optical absorption.  All of these mechanisms produce conductivity with frequency dependence: $\sigma\sim\omega^2$, but are expected to have different magnitudes.  

The first two mechanisms are specific to a U(1) spin-liquid and rely on linear coupling between the applied AC electrical field and the emergent gauge electric field. First, in a U(1) spin-liquid, an external electric field creates virtual charge fluctuations that produce emergent electric fields.  This mechanism is a strong-coupling generalization of the Ioffe-Larkin mechanism for optical absorption previously discussed in Refs.~\onlinecite{IoffeLarkin,NgLee}.  Here, we adapt strong-coupling expansion results by Bulaevskii et al.\cite{Bulaevskii}.  This approach allows us to establish a clear microscopic origin of this effect, and enables semi-quantitative estimates of its magnitude.  

Second, an external electric field can create a lattice distortion that is non-uniform within the unit cell, but uniform across different unit-cells.  Such lattice distortions perturb the spin-system with the same symmetry as the virtual polarization mechanism\cite{Bulaevskii}.  The magnitude of conductivity from the first mechanism is expected to be parametrically smaller than that of the second, since virtual charge excitations are suppressed by a factor of $\(\frac{t}{U}\)^3$ in a strong Mott insulator (where $t$ is the hopping and $U$ is the on-site repulsion).  However, it turns out that the charge-fluctuation mechanism is enhanced by a large numerical prefactor, and may actually be of the same order as the second magneto-elastic mechanism.  Both of these effects give roughly the same order of magnitude as observed in Terahertz measurements by Pilon et al.\cite{Pilon}. 

In addition, we have identified a third mechanism for absorption based on magneto-elastic and spin-orbit couplings.  Here, an applied electric field induces lattice distortions, perturbing the magnetic system.  Due to the special pattern of Dzyaloshinskii-Moriya interactions for the Kagome lattice, these spin-perturbations induce uniform spin-currents.  Absorption from this mechanism follows $\sigma\sim\omega^2\sigma_s$, where $\sigma_s$ is the spin-conductivity of the magnetic system, thereby allowing an all electrical probe of spin-transport.  Moreover, this mechanism is quite generic; it is not particular to U(1) spin-liquids, but will also produce $\sigma\sim\omega^2$ absorption in a nearly gapless $Z_2$ spin-liquid, or even in a thermal paramagnetic with diffusive spin-transport.  Absorption through this spin-orbit based mechanism is suppressed by a factor of $\(\frac{D}{J}\)^2\sim 10^{-2}$ compared to the second mechanism described above, where $D$ is the magnitude of the Dzyaloshinskii-Moriya interactions, and $J$ is the magnetic exchange coupling.

\section{Overview of $U(1)$ Dirac and $Z_2$ Spin-Liquid Scenarios}
Before describing possible sub-gap absorption mechanisms, we briefly review the slave-particle effective theory approaches for describing U(1) and $Z_2$ spin-liquids.  The ground-state properties of strong ($U\gg t$) Mott insulators at half-filling are well described by eliminating doubly occupied sites in a $t/U$ expansion to obtain an effective spin model $H_\text{Heisenberg} = J\sum_{\<ij\>}\v{S}_i\cdot\v{S}_j$, where $i$ and $j$ label sites of the Kagome lattice.  To describe quantum disordered spin-liquid states of this strongly coupled spin-theory, it is convenient to re-write the spin-degrees of freedom in terms of auxiliary Schwinger fermions or bosons: $\v{S}_i = \frac{1}{2}f^\dagger_{i,a}\boldsymbol{\sigma}_{ab}f_{i,b}$ or $\v{S}_i = \frac{1}{2}b^\dagger_{i,a}\boldsymbol{\sigma}_{ab}b_{i,b}$ respectively, subject to the constraint that $n_f=1$ ($n_b=1$).  Such descriptions  have a U(1) gauge redundancy associated with the local U(1) transformation $f_i,b_i\rightarrow e^{i\theta_i}(f_i,b_i)$, and naturally lead to low-energy effective theories with emergent U(1) gauge fields.

\subsection{U(1) Dirac Spin Liquid}
Fermionic spin-liquid states can be obtained from such spin-models by re-writing spins as Schwinger fermions: $\v{S}_i = \frac{1}{2}f^\dagger_{i,a}\boldsymbol{\sigma}_{ab}f_{i,b}$, subject to the constraint that the number of fermions, $n_f=1$.  A spin-wavefunction can be obtained by first decomposing $H_\text{Heisenberg}$ into a mean-field Hamiltonian: 
\begin{align} \label{SchwingerFermionMF} H_\text{MF} =&  J\sum_{\<ij\>}\[\chi_{ij}f^\dagger_{i,\sigma}f_{j,\sigma}+ \sum_{\<ij\>}\Delta_{ij}\(f^\dagger_{i\up}f^\dagger_{j,\down}+h.c.\)\] 
\nonumber\\
&-\mu\sum_if^\dagger_{i,\sigma}f_{i,\sigma} \end{align}
which respects the constraint $n_f=1$ only on average.  A physical spin-wave function can be obtained from the mean-field Hamiltonian by projecting out the unphysical doubly occupied sites: $|\Psi_\text{spin}\> = \mathcal{P}_G|\Psi_{H_{MF}}\>$, where $\mathcal{P}_G$ denotes a projection that removes double occupancies.  This spin-wave function can then be used as a variational ansatz. 

At the mean-field level, $\chi_{ij}^\text{MF}=\<f^\dagger_{i,\sigma}f_{j,\sigma}\>$, and $\Delta_{ij}^\text{MF} = \<f_{j,\down}f_{i,\up}\>$.  For the Kagome lattice Heisenberg model, the procedure outlined above favors a state with $\Delta_{ij}=0$ and with a sign structure of $\chi_{ij}$ such that each hexagon has an odd number of negative bonds, and each triangle has an even number\cite{RanVMC}.  One can go beyond a mean-field treatment by incorporating fluctuations in the phase of $\chi_{ij}$: $\chi_{ij}\rightarrow \chi_{ij}^\text{MF}e^{ia_{ij}}$, and in the magnitude of $\mu$: $\mu\rightarrow \mu-a^0_i$, where $a^0_i$ and $a_{ij}$ are the space- and time- components of the vector potential for the emergent U(1) gauge field.

Linearizing the low-energy spinon dispersion near the Fermi-energy, the resulting low-energy effective theory consists of four-flavors of two-component Dirac fermions (one for each combination of valley and spin), coupled to an emergent U(1) gauge field\cite{RanVMC,HermeleProperties}.   In imaginary time the effective Lagrangian density for the DSL is:
\begin{align}\label{eq:EffectiveDiracTheory} \mathcal{L} = \bar{f}_{\v{k},\omega}\[-i\omega+v_D\bs{\tau}\cdot\v{k}\]f_{\v{k},\omega}+\(j_{\v{q},\Omega}\)_\mu a^\mu_{\v{q},\Omega}\end{align}
Where we have introduced the 8-component field, $f_{\alpha}$, where $\alpha$ is a super-index labeling spin, valley, and Dirac-components, Here, $\tau$ are Pauli-matrices in the Dirac component basis\cite{HermeleProperties}, $v_D\approx J|\chi_{ij}^\text{MF}|$ is the Dirac velocity, $j = \begin{pmatrix} j^0 \\ \v{j} \end{pmatrix}$ is the spinon-number-current with $j^0_{\v{q},\Omega} =\sum_{\v{k},\omega}\bar{f}_{\v{k}-\v{q},\omega-\Omega}f_{\v{k},\omega}$ and $\v{j}_{\v{q},\Omega} = v_D\sum_{\v{k},\omega}\bar{f}_{\v{k}-\v{q},\omega-\Omega}\bs{\tau} f_{\v{k},\omega}$, and $a^\mu = \begin{pmatrix} a^0\\ \v{a}\end{pmatrix}$, where $a^0$ and $\v{a}$ are continuum versions of the long-wavelength components of the lattice-vector potentials $a^0_i$ and $a_{ij}$.

\subsection{$Z_2$ Spin-Liquids}
While the Dirac Spin-Liquid wave-function gives a good energy for the spin-1/2 Kagome Heisenberg model without any variational parameters, DMRG studies indicate that a $Z_2$ spin-liquid with a sizeable ($\approx 0.15J$) gap to spin excitations has even lower variational energy.  For Herbertsmithite, a large-gap $Z_2$ spin-liquid is inconsistent with observations of gapless behavior\cite{Helton,Imai,Olariu} down to energy scales well below $0.15J$.  However, additional ingredients such as disorder and spin-orbit coupling, not present in the pure Heisenberg model, could suppress the $Z_2$ gap, leading to a nearly gapless state consistent with experiments.  For example, slave particle mean-field studies of large $N$ analogs of the spin-1/2 Kagome Heisenberg model suggest that the spin-1/2 model is naturally close to such an $O(4)$ quantum critical point between a $Z_2$ spin-liquid and $\sqrt{3}\times\sqrt{3}$ magnetic order\cite{SachdevDM}.   

Therefore, while the main focus of this paper will be on the U(1) Dirac spin-liquid, we will also explore optical conductivity mechanisms for a nearly gapless $Z_2$ spin-liquid.  In this section, we give a brief review of the slave-particle formulation of $Z_2$ spin-liquids, and their possible phase transitions to magnetically ordered states.

$Z_2$ spin-liquids may be described in the Schwinger fermion effective field theory approach by introducing non-vanishing pairing terms $\Delta_{ij}$ into the spinon-mean-field ansatz. An alternative description can be obtained by using bosonic rather than fermionic spinons: $\v{S}_i = b^\dagger_{i,a}\frac{\bs{\sigma}_{ab}}{2}b_{i,b}$ subject to the constraint that $n_b=1$. This description also has a local U(1) redundancy leading to emergent, compact U(1) gauge degrees of freedom. In the Schwinger boson formulation, the $Z_2$ spin-liquid state is a superfluid of pairs of bosons with $\<b_{ia}\> = 0$  but $\<b_{ia}b_{ja}\>\neq 0$.  The pair condensate breaks the $U(1)$ gauge degree of freedom down to a residual $Z_2$ degree of freedom associated with $b_i\rightarrow -b_i$.  When treated exactly, the fermionic and bosonic formulations of the $Z_2$ spin-liquids are of course equivalent; and in certain cases one can explicitly construct a duality between the two descriptions, where the boson $b$ is just a vortex in the fermionic paired superfluid\cite{Ran}.

Unlike the fermionic decomposition, the bosonic formulation allows a simple description of magnetically ordered states as superfluid states with $\<b_{i,a}\>\neq 0$.  Since the frustrated Kagome geometry precludes collinear antiferromagnetic (AFM) order, it is natural to consider non-collinear (but co-planar) AFM order with $\<\v{S}(\v{r})\> = \v{n}_1\cos\v{Q}\cdot\v{r}+\v{n}_2\sin\v{Q}\cdot\v{r}$, where $\v{n_1}\cdot\v{n_2}=0$, and $\v{Q}$ is a lattice-scale wave-vector whose precise form depends on the particular type of non-collinear AFM order\cite{Chubukov}.  This formulation also naturally describes a continuous phase transition between $Z_2$ spin-liquid and non-collinear AFM, in which the spinon pairing order parameter-amplitude vanishes continuously without proliferating topological defects of the superfluid order.  The critical theory of such a transition contains gapless, deconfined spinons, and the critical properties are those of a relativistic O(4)$^*$ quantum critical point\cite{Chubukov}. Here $^*$ denotes the fractionalized critical point in which the fractionalized objects obey the ordinary O(4) critical scaling, but in which the physical degrees of freedom are composite operators of these fractionalized fields.  Interestingly, the quantum critical properties largely independent of the particular details of the type of $Z_2$ spin-liquid or non-collinear AFM order. 

\subsection{Optical Absorption - Some General Symmetry Considerations}
On symmetry grounds, a uniform external electric field, $\v{E}$, will couple to any time-reversal even and spatially uniform operator that transforms like a vector representation of the Kagome symmetry group.  In the absence of spin-orbit coupling, the electric field couples only to spin-singlet operators.  In the U(1) Dirac theory, the operators with lowest scaling dimension are the 15 fermion-bilinear mass terms\cite{HermeleProperties}:
$\v{N}_A^i = \bar{f}\tau^3\mu^i\bs{\sigma}f$, $\v{N}_B = \bar{f}\tau^3\bs{\sigma}f$, and $N_C^i=\bar{f}\tau^3\mu^if$, where $\mu^i$ are Pauli matrices in the valley basis.  Since $N_A$ and $N_C$ break translational invariance they cannot couple to a uniform $\v{E}$, and since $N_B$ is a scalar, only its spatial derivative can couple to $\v{E}$ (which would not contribute in the optical conductivity regime: $q\approx 0$, $\omega\neq 0$).  

The next lowest dimensional operators are: the emergent gauge fields $\v{e}$ and $b$, and conserved spinon-number current $\v{j}_f$ and spinon spin-current $j^i_{S^k}$ (here $i$ refers to the direction of the current of the $k$ component of spin).  In the absence of spin-orbit coupling, only $\v{e}$ has the right symmetries to couple to $\v{E}$.  In the presence of spin-orbit coupling, $\v{j}_S$ may also couple to $\v{E}$.  Therefore, we expect the electric field along the $i$-direction to couple to the operator $\mathcal{O}^i_\text{U(1)} = \lambda_1e^i+\lambda_2^{ikl}j^k_{S^l}+\dots$ where $\lambda_1$ and $\lambda_2$ are, for the moment, unknown coefficients and $\dots$ represents operators with higher numbers of time-derivatives that make negligible contributions to AC conductivity at low frequency.  As explained in more detail below, the scaling dimensions of $\v{e}$ and $\v{j}_S$ are fixed by conservation laws, indicating that the conductivity from coupling to these operators scales like $\sigma\sim\omega^2$.

Similar symmetry arguments can be applied to identify the low-energy operators of the critical $Z_2$ spin-liquid theory that may couple to an external electrical field.  While there is no gapless emergent electric field in this scenario, in the presence of spin-orbit coupling, the conserved spin-current may couple to $\v{E}$ again giving $\sigma\sim\omega^2$.  The emergent $O(4)$ symmetry of the $Z_2$-AFM quantum critical point gives rise to 9 low-dimensional operators with the small scaling dimensions\cite{Chubukov}.  If coupling to these operators were allowed, it would contribute very low power-law frequency dependence $\sigma\sim \omega^{0.5}$ based on their known scaling dimensions.  However, only higher time-derivatives of these operators can couple to $\v{E}$, giving higher power conductivity $\sigma\sim \omega^{2.5}$.  Therefore, the operator that couples to $\v{E}$ in the $Z_2$-AFM critical point is: $\mathcal{O}_{Z_2} = \lambda_3\v{j}_S+\dots$, where $\lambda_3$ is an unknown coefficient and $\dots$ represent less relevant operators.

These general considerations greatly constrain the operator form of the physical current operator, and predict conductivity scaling like $\sigma\sim \omega^2$.  In the following sections, we explore microscopic origins of these symmetry allowed couplings, in order to estimate the size of the coefficients $\lambda_{1,2,3}$.

\section{\label{sec:IoffeLarkin} Purely Electronic Mechanism for Sub-Gap Absorption in U(1) Spin-Liquids}
In states with emergent U(1) gauge fields, linear coupling between external electromagnetic fields and the emergent electromagnetic fields are generally allowed by symmetry.    Dissipation by coupling to gapless emergent degrees of freedom was originally discussed by Ioffe and Larkin\cite{IoffeLarkin} within the slave-boson approach, who argued that the physical conductivity $\sigma$ was equal to the inverse sum of the boson and spinon conductivities, $\sigma_b$ and $\sigma_f$: 
\begin{align} \label{eq:IoffeLarkin} \sigma = \frac{\sigma_b\sigma_f}{\sigma_b+\sigma_f} \end{align}
This result was later adapted specifically to the spinon-Fermi surface state and the Dirac spin-liquid by Ng and Lee\cite{NgLee}.  

Here we will revisit this absorption mechanism for the case of a Dirac spin-liquid in a strong Mott insulator.  We adapt the strong-coupling expansion results of Ref.~\onlinecite{Bulaevskii} to reveal the microscopic origin to the Ioffe-Larkin absorption mechanism, enabling us to make semi-quantitative estimates of the magnitude of this effect for Herbertsmithite.  While our results in this section are qualitatively similar to the previous study of Ref.~\onlinecite{NgLee}, we find some  differences. In particular, we find a different $t/U$ dependence of the optical absorption than obtained from a slave-rotor mean-field description.  Moreover Ref.~\onlinecite{NgLee} found that the Ioffe-Larkin mechanism produced conductivity scaling like $\sigma_\text{IL}\sim \omega^{2-\beta}$ where $\beta$ was an unknown anomalous dimension.  Here, we argue that the emergent gauge invarianes exactly fixes $\beta=0$.

\begin{figure}[ttt]
\begin{center}
\includegraphics[width = 2.25in]{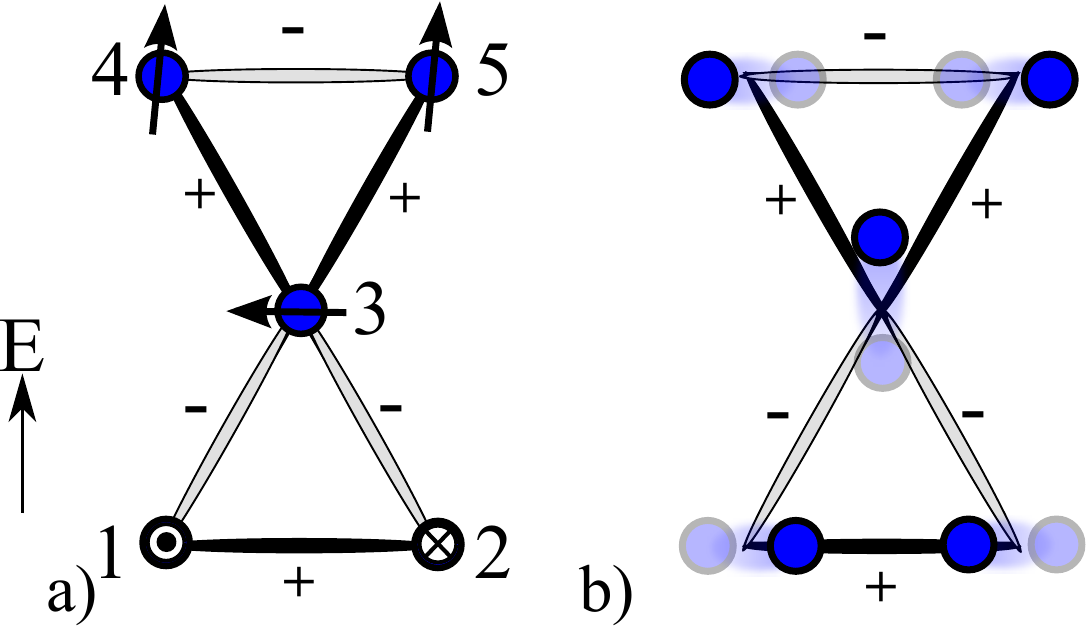}
\end{center}
\vspace{-.2in}
\caption{Panels a) and b) show the spin- and magneto-elastic mechanisms that couple the emergent gauge electric field $\v{e}$ to an applied external field $\v{E}$ shown here along the vertical (y) direction.  Panel a) depicts a spin configuration that has a net polarization along the electric field due to virtual charge fluctuations.  Dark shaded bonds labeled $+$ and light shaded bonds labeled $-$ respectively indicate strong and weak spin-singlet correlations.  Panel b) shows a distortion of the Cu$^{2+}$ ions induced by the electric field.  This distortion alters the bond distance, which increases the exchange coupling $J_{ij}$ for dark bonds labeled $+$ and decreases it for light bonds labeled $-$.  Both mechanisms have the same symmetry.}
\vspace{-.2in}
\label{fig:EeCoupling}
\end{figure}

%The Ioffe-Larkin rule, Eq.~\ref{eq:IoffeLarkin}, derives from a linear response treatment of the effective action given in Eq.~\ref{eq:SlaveRotorMFAction}.  At the level of linear response, a small external field $A$ produces a proportional response of the internal gauge field: $a_\text{ind}\sim A$ the boson and spinon currents: $j_b = K_b(a_\text{ind}+A)$, $j_f = K_fa_\text{ind}$ where $K_{b,f}^{\mu\nu}$ are the electromagnetic response kernels of the boson and fermions respectively.  By imposing the constraint $j_b+j_f=0$, required to remove extraneous non-physical states of the slave-rotor Hilbert space, one can eliminate $a_\text{ind}$ and $j_f$. Identifing $j_b$ as the physical electron current, one sees that the physical electromagnetic response kernel is given by $K = \frac{K_bK_f}{K_b+K_f}$.  In the gauge where $\v{E} = -\partial_t\v{A}$, the conductivity is simply related to the electromagnetic response kernel by $\sigma = \frac{1}{i\omega}K^{ii}$, producing Eq.~\ref{eq:IoffeLarkin}.
While real charge transitions are forbidden in the Mott insulating phase, virtual charge transitions can mediate a coupling between an externally applied physical electromagnetic field, $A^\mu$ and the emergent gapless gauge and spinon degrees of freedom of the spin-liquid, thereby generating dissipation. In p.3 of Ref.~\onlinecite{Bulaevskii}, Bulaevskii et al. derived an expression for the electron polarization and current operators for a 3-site Hubbard model triangle: $_1\hspace{-.07in}\stackrel{_3}{\triangle}\hspace{-.05in}{_2}$, in the large U limit.  The resulting operators are expressed purely in terms of spin-variables, using a strong coupling expansion in $t/U\ll 1$, to eliminate configurations with double occupancies. The leading order contribution to the polarization for a triangle of spins was found to be:
\begin{align}\label{eq:Polarization}   \v{P} &= 12ea\(\frac{t}{U}\)^3\begin{pmatrix} \v{S}_3\cdot\(\v{S}_2-\v{S}_1\)\\
\frac{1}{\sqrt{3}}\[\v{S}_3\cdot\(\v{S}_1+\v{S}_2\)-2\v{S}_1\cdot\v{S}_2\]  \end{pmatrix}
\nonumber\\
&\equiv 12ea\(\frac{t}{U}\)^3\boldsymbol{\nabla}\(\v{S}_i\cdot\v{S}_j\) \end{align} 
where $e$ is the charge of an electron (which should not be confused with $\v{e}$, the emergent electric field), $a$ is the lattice spacing (which should not be confused with the emergent vector potential $\v{a}$).  Here, $\boldsymbol{\nabla}$ is the discrete lattice derivative defined on the lattice of bond-centers with bonds labeled by their endpoints $i$ and $j$.  The x and y components of this discrete gradient are written out explicitly in the first line.

This polarization will couple the spins of each triangle linearly to an external electric field $\v{E}$, giving a perturbation to the spin-Hamiltonian:
\begin{align} \delta H_{\v{E}}(t) = -\int d^2r\v{P}(\v{r},t)\cdot\v{E}(\v{r},t)\end{align}
A small magnitude applied AC electric field $\v{E}=\v{E}_0e^{-i\omega t}$ produces an induced polarization: $\<P\>_E=\<\v{P}_{\omega}\v{P}_{-\omega}\>E(\omega)$, delivering energy through $\delta H{\v{E}}$ at a rate of $\Re e\[\<\v{P}_{\omega}\v{P}_{-\omega}\>\]E_0^2$, or, equivalently, giving conductivity $\sigma = i\omega \<\v{P}_{\omega}\v{P}_{-\omega}\>n_\triangle$ where $n_\triangle = \frac{1}{\sqrt{3}a^2}$ is the density of triangles in the kagome lattice with lattice spacing $a$.

Expressing the polarization operator $P$ as a sum of the polarizations on each triangle, re-written in terms of spin-operators through Eq.~\ref{eq:Polarization}, we find that the spin-electric field coupling produces the conductivity:
\begin{align} \label{eq:PolarizationConductivity} \sigma_\text{IL}(\omega) \approx \frac{1}{\sqrt{3}} i\omega \(12e\)^2\(\frac{t}{U}\)^6\< |\bs{\nabla}(\v{S}_i\cdot\v{S}_j)(\omega)|^2\> \end{align}

In the large $U$ limit, the bond energy gradient is directly related to the emergent electric field (see Appendix~\ref{sec:AppElectricField} for derivation):
\begin{align} \v{e} \approx \frac{J}{2}\bs{\nabla}(\v{S}_i\cdot\v{S}_j) \end{align}
where $J\approx \frac{4t^2}{U}$ is the magnetic exchange coupling.  From this expression, we see that there is a linear coupling between the physical electrical field and the emergent electromagnetic field.  Such a linear-coupling was previously identified by Ioffe and Larkin based on a slave-boson effective field theory description of the Hubbard model.  

The result presented here is simply a strong-coupling generalization of the Ioffe-Larkin result valid for $t\ll U$.  Though the functional form of the $\v{E}\cdot\v{e}$ coupling is identical to the slave-rotor treatment, the strong-coupling expansion predicts a different $t/U$ dependence compared to a mean-field slave-rotor treatment of the Hubbard model. For example, comparing to Eq.~\ref{eq:PolarizationConductivity}, and using $\sigma = \frac{1}{i\omega}\<j_\text{ph}j_\text{ph}\>$, we identify the physical electromagnetic current (in the optical conductivity limit, $\v{q}=0$, $\omega\neq 0$) as:
\begin{align} \label{eq:PhysicalEMCurrent} \v{j}_\text{ph} \approx 6ean_\triangle\frac{t}{U^2}\partial_t\v{e} \end{align}
This result disagrees with the slave-rotor mean-field theory in the large $U$ limit, which would have predicted $\v{j}_\text{ph slave-rotor} \sim \frac{t^2}{U^3}\partial_t\v{e}$ (see Appendix \ref{sec:AppSlaveRotor} for more details). We believe that this discrepancy is due to the mean-field treatment of the slave-rotor theory, which does not properly account for the gauge constraint in determining the properties of virtual charge excitations.  A simple illustration of the shortcoming of the slave-rotor mean-field, is that it predicts that gapped charge excitations propagate with velocity $ta\sim v_F$ rather than $Ja$ as would be appropriate for the strong-Mott limit (see Appendix~\ref{sec:AppSlaveRotor} for more details).  In contrast, the strong-coupling $t/U$ expansion maintains the physical electron Hilbert space, and in our view, provides a more reliable determination of the coupling between spins and external electromagnetic fields.  The results of the $t/U$ expansion can then be taken as a starting point for the slave-particle effective field theory, which we can better expect to correctly capture the low-energy physics of the spin-system.

These considerations show that the contribution to AC conductivity from  coupling to the emergent gauge field is:
\begin{align} \sigma_\text{IL}\approx \frac{72\pi}{\sqrt{3}}\frac{e^2}{h}i\omega\frac{t^2}{U^4}\<e_\omega e_{-\omega}\>\end{align}
In the absence of magnetic instantons the emergent gauge field components form a conserved current:
\begin{align} j^\mu_\text{em} &= \begin{pmatrix} b \\ \v{e}\times\hat{z}  \end{pmatrix} 
\nonumber\\
\partial_\mu j^\mu_\text{em} &= \partial_t b+\hat{z}\cdot\nabla\times\v{e}= 0\end{align}
where the second line is Faraday's law of induction for the emergent gauge field, which becomes asymptotically exact at low energies due to the emergent gauge symmetry. In a deconfined U(1) spin-liquid state with emergent scale invariance, such as the Dirac spin-liquid, the electric field cannot acquire any anomalous dimension due to this emergent conservation law.  These considerations exactly fix the scaling of $\sigma_\text{IL}\sim \omega^2$ for the Dirac spin-liquid.

To estimate the magnitude of the prefactor, we need to rely on an approximate treatment of the low-energy effective field theory.  In a random-phase approximation (RPA) treatment of the low-energy effective action, the gauge field propagator is set by the spinon electromagnetic response kernel.  Starting with the low-energy effective Dirac theory,  Eq.~\ref{eq:EffectiveDiracTheory}, choosing a gauge such that $a^0=0$, and adding a source term, $\v{J}^e$, for $\v{e}_{\v{k},\omega}=-\omega\v{a}_{\v{k},\omega}$:
\begin{align} \mathcal{L} =& \bar{f}_{\v{k},\omega}\[-i\omega+v_D\bs{\tau}\cdot\v{k}\]f_{\v{k},\omega} +\v{j}_{\v{q},\Omega}\cdot\v{a}_{\v{q},\Omega} \nonumber\\
&-\v{J}^e_{-\v{k},-\omega}\cdot\omega\v{a}_{\v{k}\omega}\end{align}
Integrating out the spinons within the RPA yields the following effective action for $\v{a}$:
\begin{align} \mathcal{L}_a = \frac{1}{2}a^i_{\v{k},\omega}K_f^{ij}(\v{k},\omega) a^j_{-\v{k},-\omega} -\v{J}^e_{-\v{k},-\omega}\cdot\omega\v{a}_{\v{k}\omega} \end{align} 
Where $K_f^{ij} = \<j^i j^j\>$ is the spinon electromagnetic response kernel.  Integrating out the emergent gauge field gives the effective action for the source $\v{J}_e$:
\begin{align} \mathcal{L}_{J^e} &= -\frac{1}{2}\omega^2\(J^e_{\v{k},\omega}\)^i \(K_f^{ij}(\v{k},\omega)\)^{-1}\(J^e_{-\v{k},-\omega}\)^j
\nonumber\\
&= \frac{\omega}{2\sigma_D} |J^e_{\v{k},\omega}|^2 \end{align}
where $\sigma_D$ is the conductivity for massless Dirac fermions (with the effective charge set to 1), which is a universal constant at low-frequency.  Analytically continuing back to real time, these manipulations show that, in the RPA, the real-time $\<ee\>$ correlator takes the form:
\begin{align} \<e^i_\omega e^j_{-\omega}\>_\text{RPA} = -\frac{i\omega}{2\sigma_D}\delta_{ij}\end{align}
The precise value of this universal constant can depend on whether one has free, interaction dominated, or disorder-dominated fermions.  Neglecting the effects of gauge-fluctuations, one has: $\sigma_D= \frac{N_f}{16} = \frac{1}{8}$.\cite{LudwigQHTransition} 

Combining this result with Eq.~\ref{eq:PolarizationConductivity} yields:
\begin{align} \label{eq:ConductivityIoffeLarkin} \sigma_\text{IL}(\omega) \approx 48\sqrt{3}\pi\(\frac{t^2\omega^2}{U^4}\)\frac{e^2}{h}\end{align}
Here the large numerical prefactor is primarily due to the factor of 12 in the relation between spin and electric polarization, Eq.~\ref{eq:Polarization}.

The experiments by Pilon et al. \cite{Pilon}, examine optical conductivity at terahertz frequencies corresponding to $\hbar\omega \approx 4$meV$\approx 50$K.  At $\omega = 1$THz, they measure three-dimensional conductivity of order $\sigma_\text{3D}(1\text{THz})\approx 0.1\Omega^{-1}\text{cm}^{-1}$.  For interlayer spacing $\sim 10\AA$, this corresponds to two-dimensional conductivity $\sigma(1\text{THz}) \approx 10^{-4}\frac{e^2}{h}$.  Comparatively, we have $U\sim 1eV$ and $t\sim U/10$.  These numbers give an estimated magnitude for the conductivity produced by coupling to the emergent electric field: $\sigma_\text{IL}(1\text{THz})\sim 10^{-5}\frac{e^2}{h}$, which is slightly smaller than, but roughly the same order as that observed.  We will see in the next section, that a coupling with the same symmetry also arises from magneto-elastic coupling, which is expected to contribute additively to this purely electronic contribution.

\section{Magneto-Elastic Mechanisms for Absorption}
In the previous section, we discussed sub-gap optical absorption in a U(1) spin-liquid through the Ioffe-Larkin coupling of the applied and emergent electric fields.  Beyond the pure spin-model, there will be additional pathways for optical conduction from magneto-elastic coupling between spin and lattice distortions and from the interplay of magneto-elastic and spin-orbit couplings.  In practice, we will argue that these mechanisms may be more important than the Ioffe-Larkin type contributions. Furthermore, in the presence of spin-orbit coupling, these additional absorption mechanisms will also be relevant for nearly gapless $Z_2$ spin-liquids without an emergent U(1) gauge field.

\subsection{Magnetoelastic Coupling to Emergent Electric Field\label{sec:MagnetoElasticCoupling}} 
Since the Cu ions have a net charge, an external electric field can give rise to a relative displacement of Cu ions within the unit cell.  As shown in Fig.~\ref{fig:EeCoupling}b, this distortion will increase the hopping strength along some bonds and decrease it along others, yielding a perturbation:
\begin{align} \label{eq:HME}\delta H_\text{ME} = \sum_{\<ij\>}\delta J_{ij}\v{S}_i\cdot\v{S}_j \end{align}
While deriving the precise distortion resulting from an electric field would require knowledge of complicated details of various lattice-elasticity parameters, it is sufficient for our purposes to simply determine what spin-operators can couple to this distortion on symmetry grounds.  The pattern of modified exchange couplings, $\delta J_{ij}$, induced by the lattice-distortion shown in Fig.~\ref{fig:EeCoupling}b can be expressed as linear combinations of the basis elements:
\begin{align} \label{eq:MagnetoElasticBasis}
\vcenter{\hbox{\includegraphics[height=18pt]{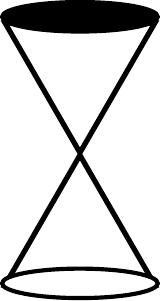}}} &\equiv \v{S}_4\cdot\v{S}_5-\v{S}_1\cdot\v{S}_2
\nonumber\\
\vcenter{\hbox{\includegraphics[height=18pt]{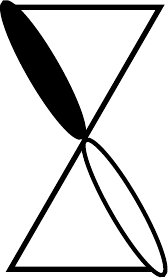}}} &\equiv 
\v{S}_4\cdot\v{S}_3-\v{S}_3\cdot\v{S}_2
\nonumber\\
\vcenter{\hbox{\includegraphics[height=18pt]{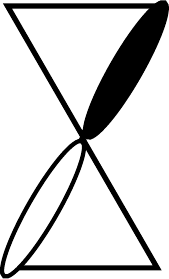}}} &\equiv \v{S}_5\cdot\v{S}_3-\v{S}_3\cdot\v{S}_1
\end{align}
Here, we represent the resulting pattern of $\delta J_{ij}$'s  pictorially for a single unit cell consisting of one upward and one downward facing triangles.  Black (white) circled-bond representing $\delta J_{ij}>0$ ($\delta J_{ij}<0$) respectively, and it is implicit that the pattern shown should be repeated uniformly throughout the Kagome lattice.

\begin{figure}[ttt]
\begin{center}
\includegraphics[width = 2in]{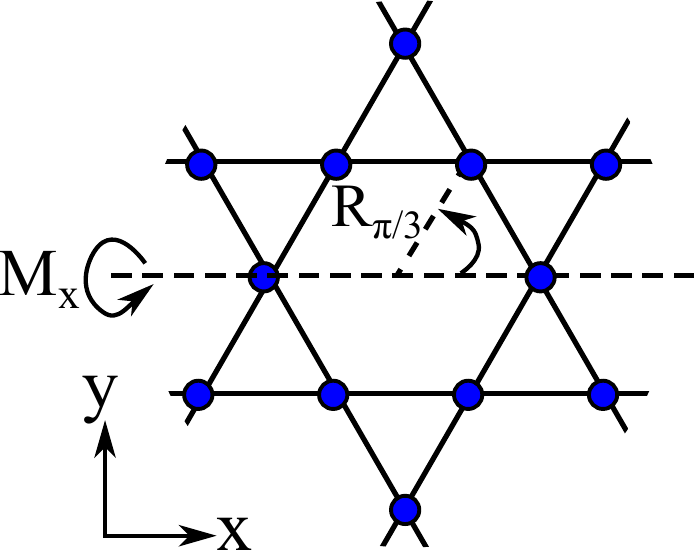}
\end{center}
\vspace{-.2in}
\caption{The Kagome point group is generated by mirror reflections, $M_x$, about the x-axis passing through the center of the vertical hour-glasses, and $\pi/3$ rotations, $R_{\pi/3}$ about the center of the hexagons.}
\vspace{-.2in}
\label{fig:KagomePointGroup}
\end{figure}

As illustrated in Fig.~\ref{fig:KagomePointGroup}, the Kagome point-group is generated by 1) clock-wise $\frac{\pi}{3}$ rotations about the center of a Hexagon, denoted $R_{\pi/3}$, and 2) mirror reflections through the horizontal axis (x-axis) passing through the points of triangles, denoted $M_x$.  The transformation properties of the basis elements in Eq.~\ref{eq:MagnetoElasticBasis} are:
\begin{align} \label{eq:MagnetoElasticTransformations}
\{
\vcenter{\hbox{\includegraphics[height=18pt]{ME1.pdf}}}\hspace{2pt},\hspace{2pt}
\vcenter{\hbox{\includegraphics[height=18pt]{ME2.pdf}}}\hspace{2pt},\hspace{2pt}
\vcenter{\hbox{\includegraphics[height=18pt]{ME3.pdf}}}
\}
&&\hspace{0.1in}
\stackrel{R_{\pi/3}}{\longrightarrow}\hspace{0.1in}&&
\{
-\vcenter{\hbox{\includegraphics[height=18pt]{ME2.pdf}}}\hspace{2pt},\hspace{2pt}
-\vcenter{\hbox{\includegraphics[height=18pt]{ME3.pdf}}}\hspace{2pt},\hspace{2pt}
-\vcenter{\hbox{\includegraphics[height=18pt]{ME1.pdf}}}
\}
\nonumber\\
\{
\vcenter{\hbox{\includegraphics[height=18pt]{ME1.pdf}}}\hspace{2pt},\hspace{2pt}
\vcenter{\hbox{\includegraphics[height=18pt]{ME2.pdf}}}\hspace{2pt},\hspace{2pt}
\vcenter{\hbox{\includegraphics[height=18pt]{ME3.pdf}}}
\}
&&\hspace{0.1in}
\stackrel{M_x}{\longrightarrow}\hspace{0.1in}&&
\{
-\vcenter{\hbox{\includegraphics[height=18pt]{ME1.pdf}}}\hspace{2pt},\hspace{2pt}
-\vcenter{\hbox{\includegraphics[height=18pt]{ME3.pdf}}}\hspace{2pt},\hspace{2pt}
-\vcenter{\hbox{\includegraphics[height=18pt]{ME2.pdf}}}
\}
\end{align}
This three dimensional representation of the Kagome point-group reduces into a one-dimensional pseudoscalar, $B_2$, irreducible representation (irrep) spanned by:
\begin{align} \mathcal{O}_{B_2} = \frac{1}{\sqrt{3}}\[
\vcenter{\hbox{\includegraphics[height=18pt]{ME1.pdf}}}
+\vcenter{\hbox{\includegraphics[height=18pt]{ME2.pdf}}}
+\vcenter{\hbox{\includegraphics[height=18pt]{ME3.pdf}}}
\] \end{align}
and a two-dimensional vector, $E_1$, irrep spanned by:
\begin{align} \label{eq:E1MagnetoElastic}
\mathcal{O}^x_{E_1} &= \frac{1}{\sqrt{2}}\[
\vcenter{\hbox{\includegraphics[height=18pt]{ME2.pdf}}}
-\vcenter{\hbox{\includegraphics[height=18pt]{ME3.pdf}}}
\]
\nonumber\\
\mathcal{O}^y_{E_1} &= \frac{1}{\sqrt{6}}\[
2\vcenter{\hbox{\includegraphics[height=18pt]{ME1.pdf}}}
-\vcenter{\hbox{\includegraphics[height=18pt]{ME2.pdf}}}
-\vcenter{\hbox{\includegraphics[height=18pt]{ME3.pdf}}}
\]
\end{align}

The external electric field $\v{E}$ is a vector, transforming like the $E_1$ irrep of the Kagome point-group, and hence only couples to the $E_1$ components: $\delta H_\text{ME} \approx \v{E}\cdot \bs{\mathcal{O}}_{E_1}$.  Comparing the symmetry of $\bs{\mathcal{O}}_{E_1}$, we see that this operator is just the discrete bond-energy gradient corresponding to the polarization operator generated described in the Eq.~\ref{eq:Polarization} of the previous section.  Therefore, as previously pointed out in Ref.~\onlinecite{Bulaevskii}, the magneto-elastic coupling due to the electric field induced displacement of Cu ions will generate a coupling of the same form as in Eqs.~\ref{eq:Polarization},\ref{eq:PolarizationConductivity}, but with a coupling constant $\frac{\partial J_{ij}}{\partial u_{ij}}\frac{\partial u_{ij}}{\partial E}\approx \frac{eJ}{K_\text{Cu}a^2}$ in place of the $\(\frac{t}{U}\)^3$ appearing in Eq.~\ref{eq:Polarization}, where $u_{ij}$ is the ionic displacement induced by the electric field, and $K_\text{Cu}$ is the effective spring constant for the Cu displacement shown in Fig.~\ref{fig:EeCoupling}b.  

For the Dirac spin-liquid, the perturbation Eq.~\ref{eq:HME} produces a linear coupling between the external and emergent electric fields, just as with the Ioffe-Larkin mechanism discussed in Sec.~\ref{sec:IoffeLarkin}.  In this case, following the derivation of Eq.~\ref{eq:ConductivityIoffeLarkin} in Sec.~\ref{sec:IoffeLarkin}, we see that this magneto-elastic coupling leads to optical conductivity:
\begin{align} \label{eq:ConductivityMagnetoElastic}\sigma_\text{ME}(\omega) \approx \(\frac{\omega}{K_\text{Cu}a^2}\)^2\frac{e^2}{h}
\end{align}
In contrast, since this magneto-elastic mechanism relies on the presence of an emergent U(1) gauge field, it is not expected to produce low-power-law frequency dependence of optical absorption for the nearly critical $Z_2$ spin-liquid scenario. The energy scale $K_\text{Cu}a^2$ is expected to be of order $\sim 1$eV,  which is comparable to $U$.  Consequently, for the Dirac spin-liquid, this absorption mechanism may dominate over the Ioffe-Larkin contribution of Eq.~\ref{eq:ConductivityIoffeLarkin}, which is smaller by a factor of $\(\frac{t}{U}\)^2$.

\subsection{Spin-Orbit Coupling}
The mechanisms described in the previous section relied on the existence of an emergent electric field, a low-dimensional operator with the right symmetries to couple to the external electric field (i.e. a time-reversal invariant, spatially uniform, spin-singlet operator transforming under the $E_1$ representation of the Kagome point group).  

These mechanisms are special to the U(1) spin-liquid, since, in the alternate scenario where the ground state is a nearly critical $Z_2$ spin-liquid, there are no spin-singlet operators with the right symmetry properties.
However, as we will show in this section, the presence of spin-orbit coupling enables the external electric field to couple not only to spin-singlet operators, but also time-reversal invariant spin-pseudo-vectors.  In particular, the combination of spin-orbit and magneto-elastic couplings will generically enable the electric field to couple to spin-current, enabling purely electrical probes of spin-transport.  This coupling to spin-current provides alternative mechanisms for optical absorption which lead to AC electrical conductivity scaling like $\sigma\sim \omega^2$ in any gapless relativistic state.  In particular this mechanism is also relevant in a nearly gapless state, at finite frequency above the spin-gap, for example in the vicinity of an O(4) quantum critical point between a $Z_2$ spin-liquid and non-collinear antiferromagnet.

\subsubsection{Dzyaloshinskii-Moriya (DM) Terms}
We begin with a brief overview of spin-orbit coupling in Herbertsmithite and related Cu spin-1/2 Kagome antiferromagnets.  Due to the atomic spin orbit coupling in the Cu d-orbitals, $\lambda_\text{so}$, and because spin-exchange occurs along a bent Cu-O-Cu super-exchange pathways, Herbertsmithite has substantial Dzyaloshinskii-Moriya (DM) interactions: $\sum_{\<ij\>}\v{D}_{ij}\cdot\(\v{S}_i\times\v{S}_j\)$, where the DM vector $D\approx \frac{\lambda_\text{so}t}{U}\sin\alpha$ where $\alpha$ is the angle of the Cu-O-Cu oxygen bond (defined such that $\alpha=0$ for a linear super-exchange pathway).  For Herbertsmithite, the DM interactions are expected to be roughly $10\%$ of the exchange energy: $D\approx 0.1J$.  

The dominant effects of the DM interactions come from the z-component of the DM vectors, $\hat{z}\cdot\v{D}_{ij}$, where the $\hat{z}$ denotes the c-axis perpendicular to the Kagome planes.  As we will see below, the DM terms generate coupling between external electric fields and spin-current.  Since the in-plane components sum to zero within the unit cell, these components can only create a response to spatially inhomogeneous electric fields that vary in sign over short distances of order a single lattice spacing.  Such sharp spatial variations are not relevant for AC conductivity measurements that probe the response to long wave-length fields, and therefore the in-plane DM components may be neglected for what follows.
% (formally, one can view the spin-orbit terms as a non-dynamical SU(2) gauge field for the electrons, and the vanishing of the sum of in-plane DM terms within the unit cell allows one ``gauge" away the in-plane DM terms to $\mathcal{O}\(\frac{D}{J}\)^2$ by local spin rotations). The pattern of $\hat{z}\cdot\v{D}_{ij}$ for Herbertsmithite is shown in Fig.~\ref{fig:DMModulation}a).  

In a Schwinger-fermion mean-field description of the Heisenberg model, the DM terms simply become spin-orbit hopping terms for the spinons:
\begin{align} \label{eq:DMMF} \v{D}_{ij}\cdot\(\v{S}_i\times\v{S}_j\)\rightarrow i\frac{\chi^f_{ij}}{J}\v{D}_{ij}\cdot\(f^\dagger_{i,a}\boldsymbol{\sigma}_{ab} f_{j,b}\) \end{align}
where $\chi^f = \<f^\dagger_{i\sigma}f_{j\sigma}\>$.  
For the Dirac spin-liquid state, this pattern of DM interactions translates into a quantum spin-Hall mass for the Dirac spinons.  Such a term is expected to gap out the Dirac spinons, enabling gauge-instantons to proliferate.  Due to the quantum spin-Hall response, such instanton flux-insertion events are bound to spin-flips, and their proliferation leads to magnetic order\cite{HermeleProperties}.  Similarly, Schwinger-boson mean-field treatments of the Kagome-Heisenberg model indicate that the $Z_2$ spin-liquid state for the bare Heisenberg model is destabilized by finite DM interactions, which are predicted to drive the model to non-collinear antiferromagnetic order\cite{SachdevDM}. Experimentally, Herbertsmithite appears to avoid magnetic order to the lowest achievable temperatures, and is either gapless or has a very small gap.  

\begin{figure}[ttt]
\begin{center}
\includegraphics[width = 2.5in]{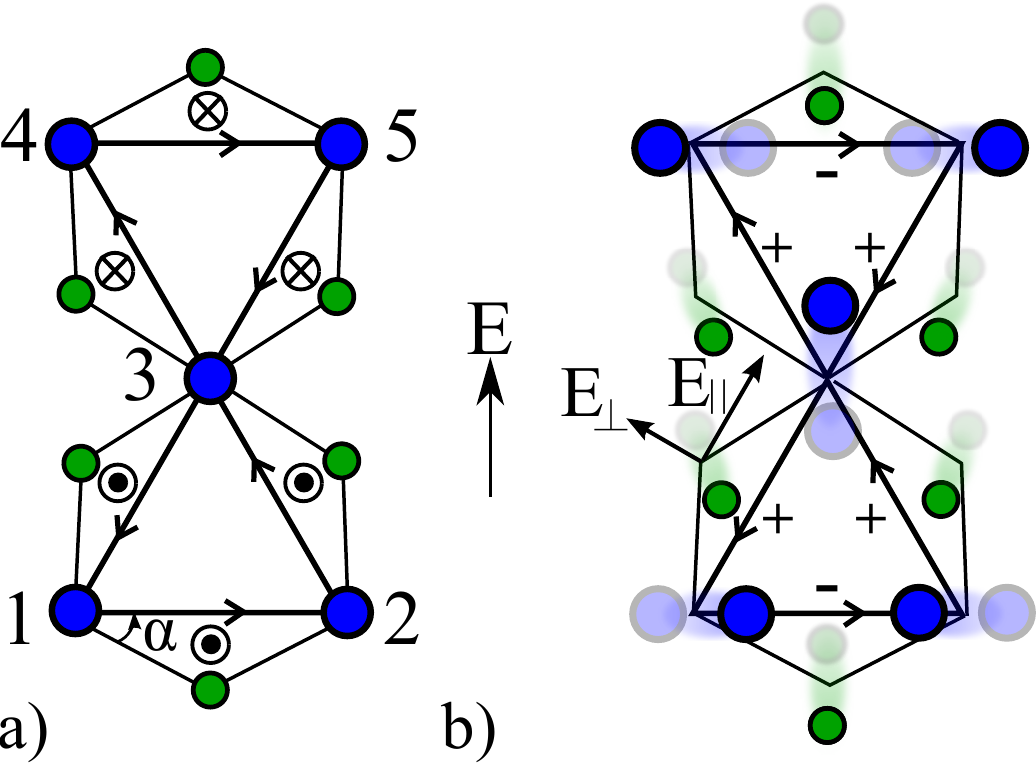}
\end{center}
\vspace{-.2in}
\caption{Blue and green dots are Cu and O ions respectively.  a) DM arises from the bent Cu-O-Cu bonds.  The pattern of $\v{D}$ vectors is shown by $\otimes$ and $\odot$ indicating $\v{D}$ into and out of the plane respectively for the bond-orientation shown with arrows.  b) In an electric field, the oppositely charged Cu$^{2+}$ and O$^{2-}$ ions undergo relative displacements.  The O ion displacement changes the bond-angle, and the Cu displacement changes the bond exchange strength as described in Fig.~\ref{fig:EeCoupling}.  As explained in the text, this alters the DM pattern as indicated by the $+$ and $-$ labels, which couples the electric field to the spin-current.}
\vspace{-.2in}
\label{fig:DMModulation}
\end{figure}

\subsubsection{DM Modulation}
The lattice distortion discussed above in Sec.~\ref{sec:MagnetoElasticCoupling} alters the Cu to Cu bond lengths following the pattern shown in Eq.~\ref{eq:E1MagnetoElastic}, thereby changing the hopping strengths along these bonds in the same manner.  Since the DM vector on bond $ij$ is related to the Cu-Cu hopping strength by: $D_{ij}\sim \frac{t_{ij}\lambda_\text{so}}{U}$, this distortion also changes the magnitude of the DM vectors in the same manner, introducing the perturbation: 
\begin{align} \delta H_\text{DM}^{\text{Cu Distortion}} = \sum_{\<ij\>} \alpha_{ij}\hat{z}\cdot\(\v{S}_i\times \v{S}_j\)
\end{align}

For definiteness, we consider $E$ along the vertical (y-) direction as shown in Fig.~\ref{fig:DMModulation}b.  In this case, the magnitudes, $|\alpha_{ij}|$, are twice as large on the horizontal bonds ($\overline{12}$ and $\overline{45}$) as on the angled bonds ($\overline{23}$, $\overline{31}$, $\overline{34}$ and $\overline{53}$). The sign structure of $\alpha_{ij}$ can be represented pictorially by:
\begin{align}\label{eq:HDM} \delta H_\text{DM} \sim  2\vcenter{\hbox{\includegraphics[height=24pt]{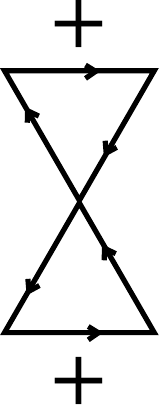}}}
-\vcenter{\hbox{\includegraphics[height=18pt]{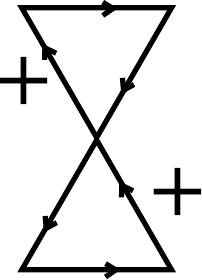}}}
-\vcenter{\hbox{\includegraphics[height=18pt]{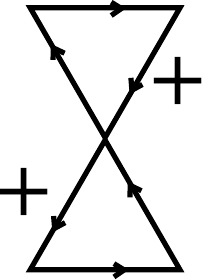}}}\end{align}
where we have expanded the coupling as a linear combination of the basis operators:
\begin{align} \label{eq:OxyModePrimitives}
\vcenter{\hbox{\includegraphics[height=24pt]{Oxy1.pdf}}} &\equiv \hat{z}\cdot\(\v{S}_1\times\v{S}_2\)+\hat{z}\cdot\(\v{S}_4\times\v{S}_5\)
\nonumber\\
\vcenter{\hbox{\includegraphics[height=18pt]{Oxy2.pdf}}} &\equiv \hat{z}\cdot\(\v{S}_3\times\v{S}_2\)+\hat{z}\cdot\(\v{S}_4\times\v{S}_3\)
\nonumber\\
\vcenter{\hbox{\includegraphics[height=18pt]{Oxy3.pdf}}} &\equiv \hat{z}\cdot\(\v{S}_5\times\v{S}_3\)+\hat{z}\cdot\(\v{S}_3\times\v{S}_1\)
\end{align}
which form a closed set under the action of the Kagome point group.

Since the oxygen and Cu ions are oppositely charged, they will generally they will undergo a relative displacement in a uniform electric field, $E$.  In particular, the oxygen bond angles will be changed thereby changing the DM terms as indicated in Fig.~\ref{fig:DMModulation}b.  To linear order the DM change is affected only by the bond-angle change, which depends only to the component oxygen displacement perpendicular to the local bond direction within the kagome plane.  For definiteness, we consider $E$ along the vertical direction in Fig.~\ref{fig:DMModulation} resolving the applied field in to directions along and perpendicular to the local bonds: $E_{ij,\parallel}$  and  $E_{ij,\perp}$, as shown in Fig.~\ref{fig:DMModulation}b.  It can be readily seen that this contributes a term of the same form as Eq.~\ref{eq:HDM}.

This perturbation, $\delta H_\text{DM}^{\text{O}^{2-}}$, can couple to any spin-operator with the same symmetries.  Therefore, it is instructive to determine its transformation properties of under the Kagome lattice point-group by considering irreducible representations constructed from the basis of Eq.~\ref{eq:OxyModePrimitives}.  In the presence of spin-orbit coupling terms, the Kagome point-group remains unchanged but the spatial point-group transformations must also be accompanied by spin-rotations\cite{HermeleProperties}.  Since, as explained above, we may neglect the in-plane components of the DM vectors the only modification is that mirror reflections about the x-axis must be accompanied by $\pi$ spin-rotations around the x-axis.

Under the action of $R_{\pi/3}$ and $M_x$, the vector chirality combinations of Eq.~\ref{eq:OxyModePrimitives} transform like: 
\begin{align} \{\vcenter{\hbox{\includegraphics[height=24pt]{Oxy1.pdf}}}
\hspace{2pt},\hspace{2pt}
\vcenter{\hbox{\includegraphics[height=18pt]{Oxy2.pdf}}}
\hspace{2pt},\hspace{2pt}
\vcenter{\hbox{\includegraphics[height=18pt]{Oxy3.pdf}}}\}
&&\hspace{0.1in}
\stackrel{R_{\pi/3}}{\longrightarrow}\hspace{0.1in}&&
-\{\vcenter{\hbox{\includegraphics[height=18pt]{Oxy2.pdf}}}
\hspace{2pt},\hspace{2pt}
\vcenter{\hbox{\includegraphics[height=18pt]{Oxy3.pdf}}}
\hspace{2pt},\hspace{2pt}
\vcenter{\hbox{\includegraphics[height=24pt]{Oxy1.pdf}}}\}
\nonumber
\\
\{\vcenter{\hbox{\includegraphics[height=24pt]{Oxy1.pdf}}}
\hspace{2pt},\hspace{2pt}
\vcenter{\hbox{\includegraphics[height=18pt]{Oxy2.pdf}}}
\hspace{2pt},\hspace{2pt}
\vcenter{\hbox{\includegraphics[height=18pt]{Oxy3.pdf}}}\}
&&\hspace{0.1in}
\stackrel{M_x}{\longrightarrow}\hspace{0.1in}&&
-\{
\vcenter{\hbox{\includegraphics[height=24pt]{Oxy1.pdf}}}
\hspace{2pt},\hspace{2pt}
\vcenter{\hbox{\includegraphics[height=18pt]{Oxy3.pdf}}}
\hspace{2pt},\hspace{2pt}
\vcenter{\hbox{\includegraphics[height=18pt]{Oxy2.pdf}}}
\}
\end{align}
where the minus sign in the first line arises because rotations invert our chosen bond-orientation, swapping $\v{S}_i\times\v{S}_j\rightarrow \v{S}_j\times\v{S}_i = -\v{S}_i\times\v{S}_j$.  Similarly the minus sign in the second comes from the extra spin-rotation that accompanies the mirror reflection $M_x$.  Furthermore, since all of these combinations arise from linear combinations of vector spin-chirality, each of the basis vectors are invariant under time-reversal symmetry.

The three dimensional representation of the Kagome point-group spanned by the basis shown in Eq.~\ref{eq:OxyModePrimitives} is reducible, and decomposes into one-dimensional pseudo-scalar, A2,  irreducible representation (irrep) spanned by (see Appendix C of Ref.~\onlinecite{HermeleProperties} for more details about the representation theory for the Kagome point-group):
\begin{align} O_{\text{DM},A_2} = \frac{1}{\sqrt{3}}\[\vcenter{\hbox{\includegraphics[height=24pt]{Oxy1.pdf}}} +\vcenter{\hbox{\includegraphics[height=18pt]{Oxy2.pdf}}} +\vcenter{\hbox{\includegraphics[height=18pt]{Oxy3.pdf}}}\] \end{align}
and a two-dimensional vector, $E_1$, irrep spanned by:
\begin{align} O_{\text{DM},E_1}^y &= \frac{1}{\sqrt{6}}\[2\vcenter{\hbox{\includegraphics[height=24pt]{Oxy1.pdf}}} -\vcenter{\hbox{\includegraphics[height=18pt]{Oxy2.pdf}}} -\vcenter{\hbox{\includegraphics[height=18pt]{Oxy3.pdf}}}\] 
\nonumber\\ 
O_{\text{DM},E_1}^x &= \frac{1}{\sqrt{2}}\[\vcenter{\hbox{\includegraphics[height=18pt]{Oxy2.pdf}}} -\vcenter{\hbox{\includegraphics[height=18pt]{Oxy3.pdf}}}\] \end{align}

The Cu bond-, and oxygen bond angle- distortions couple only to the $E_1$ vector piece: $\delta H_\text{DM}\sim \v{E}\cdot \v{O}_{v}$.  To summarize, $\sum_{\<ij\>} \alpha_{ij}\hat{z}\cdot\(\v{S}_i\times \v{S}_j\)$ is time-reversal invariant, transforms as the z-component of a vector under spin-rotations, and transforms like a $E_1$ vector representation of the Kagome point-group.  These transformation properties are exactly the same as the z-component of the spin-current,  $\v{j}_{S^z}$.  Therefore, there one should generically expect a linear coupling between the electric field and the spin-current, mediated by the oxygen bond distortion shown in Fig.~\ref{fig:DMModulation}b.  Interestingly, this coupling implies that one can directly probe spin-conductivity of Kagome materials with the DM pattern shown in Fig.~\ref{fig:DMModulation}a, purely through electrical measurements.

This expectation can be explicitly verified. Noting that the spin-current operator obeys the continuity equation $\frac{\partial S^a}{\partial t} = -i[H,S^a]= -\nabla\cdot\v{J}_{S^a}$, and using the Heisenberg Hamiltonian to compute the time-evolution, one finds that the a-component of spin-current along the $\overrightarrow{ij}$ bond is just $\(\v{S}_j\times\v{S}_i\)^a$. Alternatively, for the spin-liquid scenarios, one can perform a mean-field decomposition of the perturbation term within the Schwinger fermion (or boson) framework using Eq.~\ref{eq:DMMF}.  In either case, one finds that the perturbation, $\delta H_\text{DM}\sim \v{E}\cdot \v{O}_{v}$, induces a uniform spin current, $\v{j}_{S^z}$, perpendicular to the applied $\v{E}$ field in the $\hat{z}\times\v{E}$ direction.  Hence, the optical conductivity resulting from this mechanism will be proportional to $\omega^2$ times the spin-conductivity.  

Modeling the Cu and Oxygen bond-angle distortions as springs with  effective spring-constant $K_\text{Cu}$ and $K_\text{O}$ respectively, one can estimate the magnitude of the induced coupling:
\begin{align} \label{eq:deltaHDM} \delta H_\text{DM} &\approx  \frac{2eD\chi^f}{JK_\text{eff}a}\v{E}\cdot\(\hat{z}\times \v{j}_{S^z}\) 
\end{align}
where $K_\text{eff} = \(\frac{1}{K_\text{O}\cos\alpha}-\frac{1}{K_\text{Cu}}\)^{-1}$ is the effective spring constant accounting for the fact that the Cu$^{2+}$ and O$^{2-}$ ions have opposite charge, and $\alpha$ is the Cu-O bond-angle.  

For the Dirac spin-liquid, Eq.~\ref{eq:deltaHDM} indicates that an external electric field induces a spin-current $j_{S^z}$:
\begin{align} \v{j}_{S^z} \approx \frac{D}{J}\frac{1}{K_\text{eff}a}\(i\omega\sigma_D\)\(\hat{z}\times\v{E}\) \end{align}
where $\sigma_D$ is the spinon-conductivity.  Similar results will hold for the $\mathbb{Z}_2$ case.

Since $\v{j}_{S^z}$ is a conserved quantity, its scaling dimension is fixed in any scale invariant theory, like the Dirac spin-liquid or in the vicinity of a quantum-critical point, resulting in $\sigma(\omega)\sim \omega^2$. 
\begin{align}\label{eq:ConductivityDM}\sigma_\text{DM}\approx \frac{e^2}{h}\(\frac{D}{J}\)^2\(\frac{\omega}{K_\text{eff}a^2}\)^2\sigma_s \end{align}
where $\sigma_s = \frac{1}{i\omega}\<j_{s^z}j_{s^z}\>$ is the spin-conductivity.

For the Dirac spin-liquid: $\sigma_s\approx \sigma_D$ is constant. Comparing to the contribution from the spin-symmetric magneto-elastic coupling of Eq.~\ref{eq:ConductivityMagnetoElastic}, we see that the magnitude of the Ioffe-Larkin mechanism may dominate due to the factor of $\(\frac{D}{J}\)^2\approx 10^{-2}$.  However, for the scenario of a nearly gapless $Z_2$ spin-liquid, the spin-conductivity coupling may be the dominant absorption mechanism.

\subsubsection{Spin-Orbit Mechanism in a Thermal Paramagnet}
While we have illustrated this result explicitly for the U(1) Dirac spin-liquid, the DM mediated coupling between electric field and spin-current is generically allowed by symmetry, and will be present in any theory of Kagome spin systems with DM interactions.  Furthermore, for any scale invariant theory, the resulting conductivity from coupling to this phonon mode will scale as $\sim \omega^2$, including a Dirac spin-liquid or any nearly gapless $Z_2$ spin-liquid in the quantum critical regime near a continuous phase transition, for example to non-collinear antiferromagnetic order.

Furthermore, the coupling of an electric field to spin-current may be the most important mechanism for thermal paramagnetic states at intermediate to high temperature ranges, well above the low-energy scale $\Delta_S$.  For example, the low energy scale $\Delta_S$ would be the spin-gap of a $Z_2$ spin-liquid, the magnetic ordering scale of the frustrated antiferromagnet, or the spinon bandwidth in a $U(1)$ spin-liquid (note that this scale is well-separated from $J$ in a frustrated lattice like the Kagome).   Regardless of the nature of the spin ground-state at $T\ll \Delta_S$, for $T>\Delta_S$ one can expect diffusive spin-transport with non-vanishing spin-diffusion coefficient $D_s$, which tends to $D_s\approx Ja^2$ at high temperature\cite{Lurie}. $D_s$ is related to spin-conductivity, $\sigma_s$, by the Einstein relation $\sigma_s\approx D_s\chi_s$ where $\chi_s$ is the spin-susceptibility. At intermediate to high temperature one expects Curie-Weiss-like spin-susceptibility: $\chi_s(\omega,T)\approx \frac{1}{T+\theta_\text{CW}}$ with $\theta_\text{CW}\approx J>0$. In Herbertsmithite, for example, $\chi_s$ is a slowly increasing function of $T$ down to $T\approx 50$K\cite{Helton}, giving rise to a constant spin-conductivity, and power-law electrical conductivity $\sigma\sim\omega^2$. We note that this effect may be difficult to observe in practice, even at higher temperatures, due to the suppression of this conductivity-pathway by a factor of $\(\frac{D}{J}\)^2$.

\section{Discussion}
To summarize, we have found two different mechanisms for sub-gap absorption in a spin-liquid Mott insulator on the Kagome lattice.  For a U(1) spin-liquid, absorption can occur through coupling the external electric field to the emergent electric field mediated either by virtual charge fluctuations or magneto-elastic effects.  This mechanism produces $\sigma\sim \omega^2$ conductivity for the Dirac spin-liquid.  

Second, the interplay of spin-orbit and magneto-elastic couplings creates a direct coupling of an electric field to the spin-current of the Kagome system.  Therefore, optical conductivity measurements are also probing spin-conductivity.  This mechanism is quite generic, and will produce $\sigma \sim \omega^2$ electrical conductivity in any gapless relativistic phase or critical point, including the Dirac spin-liquid and the vicinity of the O(4) deconfined critical point between a $Z_2$ spin-liquid and a non-collinear magnetically ordered state.  

In contrast, this mechanism will not lead to substantial power-law conductivity deep inside a magnetically ordered state.  A magnetically ordered state can be regarded as a spin-superfluid, with spin-waves corresponding to the superfluid density mode.  In this state, the spin-current is given by the spatial gradient of the superfluid phase, and due to the huge mismatch in the velocity of light, c, and the spin-wave velocity, $Ja$, an electromagnetic field can only couple to multi-spin-wave emission processes giving a negligible contribution.

Therefore, at asymptotically low frequency and temperature, power-law optical conductivity can likely be taken as a sign of gapless (or nearly gapless) spinon excitations.  At intermediate to high temperatures where the spin-system is a thermal paramagnet, the coupling to spin-current will produce $\sigma\sim \omega^2$ due to spin-diffusion, regardless of the nature of the low-temperature spin-phase (its magnitude may be small due to the factor of $\(\frac{D}{J}\)^2$ in Eq.~\ref{eq:ConductivityDM}).  However, at temperatures below $\sim 50K$ the spin-susceptibility $\chi_s$ dips sharply with decreasing $T$,\cite{Olariu} indicating the onset of correlations and a departure from the trivial thermal paramagnet.

%The temperature dependence of conductivity can possibly be used to distinguish such a scenario.  For example, the thermal paramagnet will have conductivity $\sigma\sim \omega^2D_s\chi_s(T)$. If we assume that the high-temperature spin diffusivity $D_s$ is roughly $Ja^2$ independent of temperature, then absorption through thermal spin-diffusion should decrease in magnitude with decreasing temperature (for temperatures lower than $\sim 50K$, below which the spin susceptibility $\chi_s$ is a decreasing function of $T$)\cite{Olariu}.  In contrast, the experimental observations of Pilon et al. show the opposite temperature dependence for the low-power frequency dependence of $\sigma$: namely the power-law component increases gradually with decreasing temperature\cite{Pilon}.

Within the two spin-liquid scenarios, the temperature dependence of the coefficient of the $\sim\omega^2$ optical absorption is more subtle.  Within the scaling regime for both theories, the optical conductivity will act like $\omega^2$ times a universal function of $\frac{\omega}{T}$.  For the nearly critical $Z_2$ spin-liquid, the dominant mechanism is proportional to the spin-conductivity.  For the O(4)$^*$ critical point, in the high-frequency ($\omega\gg T$) regime relevant to the experiments of Pilon et al.\cite{Pilon}, the conductivity is expected to gradually increase, saturating to a constant value as $\frac{\omega}{T}\rightarrow \infty$ (see e.g. Ref.~\onlinecite{SachdevBook}).  This temperature dependence is in agreement with that observed in Ref.~\onlinecite{Pilon}.

As described above, in the Dirac spin-liquid case, the Ioffe-Larkin type mechanisms dominate.  The resulting conductivity is proportional to the spinon-resistivity $\rho_D(\frac{\omega}{T}) = \(\sigma_D(\frac{\omega}{T})\)^{-1}$.  Here the finite temperature and frequency scaling is known only in the limit of large-number of Dirac flavors, $N_f$.  For $N_f\rightarrow \infty$, one recovers the case of free Dirac fermions for which $\rho_D$ decreases with decreasing temperature in the $\omega\gg T$ limit.  This $N_f=\infty$ prediction shows the opposite temperature dependence of that observed in Pilon et al.'s experiment\cite{Pilon}.  However we caution that it is unclear to what extent this prediction applies to the physical case of $N_f=2$, given the absence of a controlled theory in this limit.  For example, in other strongly coupled field theories based on holographic duality the temperature dependence at high frequency can be of arbitrary sign depending on details\cite{SachdevHolography}.

\textit{Acknowledgements} - We thank Naoto Nagaosa for discussions leading to the derivation in App.~\ref{sec:AppElectricField}, as well as helpful discussion from Nuh Gedik, Daniel Pilon, Maksym Serbyn, and Oleg Starykh.

ACP and PAL acknowledge funding for this research by: NSF grant DMR 1104498. In addition ACP and PAL thank the Kavli Institute for Theoretical Physics at UC Santa Barbara, for hosting part of this research, supported in part by the National Science Foundation under Grant No. NSF PHY11-25915. TS was supported by NSF DMR-1005434, and also partially supported by the Simons Foundation by award number 229736. He thanks the Physics Department at Harvard for hospitality where part of this work was done. 

\appendix

\section{Microscopic Identification of the Emergent Electromagnetic Fields \label{sec:AppElectricField}}
The emergent gauge degrees of freedom arising in the slave-rotor effective theory are sometimes described as ``fictitious", as they arise out of a redundancy of the slave-rotor formalism.  However, while the emergent vector potential, $a^\mu$, has no direct physical meaning, gauge invariant quantities derived from $a^\mu$ have observable physical meaning in terms of spin.  

The functional form of the expression of the emergent electric and magnetic fields may be deduced on symmetry grounds and for a strong Mott insulator.  Furthermore, their parametric dependence on the parameters of the Hamiltonian can be fixed by dimensional analysis, at least for a strong Mott insulator where there is only one length scale, $a$, and one energy scale, $J$.  It is nevertheless gratifying to produce an explicit derivation of the microscopic meaning of the gauge flux through an elementary plaquette and electric field along a lattice link.  One should keep in mind, however, that these microscopic operators will be renormalized in going to a low-energy effective description and will receive contributions from all other low-energy operators with the same symmetries.

By considering spatial Wilson loops of $a^\mu$ within the slave-particle effective theory, Refs.~\onlinecite{LeeNagaosa,WenWilczekZee} demonstrated that the emergent magnetic flux through a triangular plaquette corresponds to scalar spin-chirality of the triangle:
\begin{align} b_{ijk} \sim \v{S}_i\cdot\(\v{S}_j\times\v{S}_k\)\end{align}
where sites $i,j,k$ form corners of the triangle (traversing cyclically in the clock-wise direction).  As shown in Refs.~\onlinecite{Motrunich,Bulaevskii}, in a strong Mott insulator, the vector spin-chirality is just the electrical current circulating around the triangle in the clock-wise sense: 
\begin{align} b_{ijk}\sim j^\text{circ}_{ijk} \approx -e\(\frac{24t^3a}{U^2}\)\v{S}_i\cdot\(\v{S}_j\times\v{S}_k\) +\mathcal{O}\(\frac{t^5}{U^4}\) \end{align}

In this appendix, we follow a similar route to establish the physical meaning of the emergent electrical field $\v{e}$.  We will show that, in terms of the physical spin-degrees of freedom, the electric field corresponds to the gradient of bond-energy, which in turn corresponds to the electrical polarization arising from virtual charge fluctuations (as described in Ref.~\onlinecite{Bulaevskii}).

\subsection{Spatial Wilson Loops and the Emergent Magnetic Field}
We begin by reviewing the arguments of Ref.~\onlinecite{LeeNagaosa} to identify $b$ as the scalar spin-chirality.  In the Schwinger-fermion approach one rewrites spins as auxiliary fermions: $\v{S}_i = f^\dagger_{i,a}\frac{\boldsymbol{\sigma}_{ab}}{2}f_{i,b}$, subject to the constraint $\sum_{\sigma}f^\dagger_{i,\sigma}f_{i,\sigma}=1$. Converting the resulting Heisenberg Hamiltonian into a functional integral with fields $\bar{f},f$ (corresponding to $f^\dagger,f$) and introducing Hubbard-Stratonovich fields $\chi_{ij}$ to decouple the resulting four-fermion hopping term yields the following term in the effective action:
\begin{align} \mathcal{L}_\text{J}=&
J\bar{f}_{i,a}(\tau)f_{j,a}(\tau)\bar{f}_{j,b}(\tau)f_{i,b}(\tau)\nonumber\\
& \longrightarrow
J\chi_{ij}\bar{f}_{i,a}f_{j,a}+J\bar{\chi}_{ij}\bar{f}_{j,a}f_{i,a}
+\frac{J}{2}\bar\chi_{ij}\chi_{ij}
\end{align}
In the resulting low-energy theory, fluctuations of the magnitude of $\chi_{ij}$, around their saddle-point (mean-field) value are massive, and the phase fluctuations about the saddle point value are simply $e^{ia_{ij}}$.  The Wilson loop around a spatial loop $\Gamma$ is: 
\begin{align} W_\Gamma = \prod_{\circlearrowleft\Gamma}e^{ia_{ij}} \approx \prod_{\circlearrowleft\Gamma}\chi_{ij} \end{align}
where $a_{ij}$ is the vector-potential on the bond $\<ij\>$.  

This Wilson loop can be generated by introducing an infinitesimal source term 
\begin{align}\delta\mathcal{L}=\bar{\eta}_{ij}\chi_{ij}+\eta_{ij}\bar{\chi}_{ij} \end{align}
By making the change of variables:
\begin{align} \chi_{ij}\rightarrow \chi_{ij}-\eta_{ij}\hspace{0.2in} \bar\chi_{ij}\rightarrow \bar\chi_{ij}-\bar\eta_{ij} \end{align}
one shifts the source for $\chi_{ij}$ to a source for $f^\dagger_{i,a}f_{j,a}$, indicating that the Wilson loop is equivalently expressed in terms of a string of fermion fields:
\begin{align} W_\Gamma = \prod_{\circlearrowleft\Gamma}e^{ia_{ij}} \approx \prod_{\circlearrowleft\Gamma}\bar{f}_{i,a}f_{j,a} \end{align}
For example, the flux through a triangular plaquette is $W\hspace{-.07in}{{_k}\atop{_i\triangle_j}}\bar{f}_{i,a}f_{j,a}\bar{f}_{j,b}f_{k,b}\bar{f}_{k,c}f_{i,c}$.

One can then restore the Grassman expression to operator form\cite{OperatorOrderingEndNote}: $W\hspace{-.07in}{{_k}\atop{_i\triangle_j}} \rightarrow f^\dagger_{i,a}f_{j,a}f^\dagger_{j,b}f_{k,b}f^\dagger_{k,c}f_{i,c}$
which can be expressed purely in terms of physical variables of through the identities:
\begin{align} f^\dagger_{i,a}f_{i,b} &= \frac{\delta_{ab}}{2}n_{f,i}+\v{S}_i\cdot\boldsymbol{\sigma}_{ab}\approx \frac{\delta_{ab}}{2}+\v{S}_i\cdot\boldsymbol{\sigma}_{ab}
\nonumber\\
f_{i,a}f^\dagger_{i,b} &= \frac{\delta_{ab}}{2}(2-n_{f,i})-\v{S}_i\cdot\boldsymbol{\sigma}_{ab}
\approx \frac{\delta_{ab}}{2}-\v{S}_i\cdot\boldsymbol{\sigma}_{ab}
\end{align}
where $n_{f,i}$ is the spinon occupation number of cite $i$, which can be regarded as approximately constant: $n_f\approx 1$, and $\sigma$ are the spin-1/2 Pauli matrices.

The magnetic flux through a triangle $ijk$ is then $b_{ijk}=\frac{1}{2i}\(W_{ijk}-W_{kji}\)=\Im m \(W_{ijk}\) \approx \v{S}_i\cdot\(\v{S}_j\times\v{S}_k\)$.  The same procedure identifies the magnetic flux through a square plaquette with vertices $ijkl$ as:
\begin{align} b_{ijkl} = \frac{1}{2}\(b_{ijk}+b_{jkl}+b_{kli}+b_{lij}\)\end{align}

\subsection{Emergent Electric Field}
Proceeding by analogy to the previous section, the electric field on a link $\<ij\>$ is given by the space-time Wilson loop:
\begin{align}\label{eq:SpaceTimeWilsonLoop}e_{ij} &= \lim_{\delta t\rightarrow 0}\frac{1}{\delta t} \[e^{i\(a_{ij}(t)+a^0_j(t)-a_{ij}(t+\delta t)-a^0_i(t)\)}\]
\nonumber\\
&\rightarrow \frac{1}{2}\[\v{S}_i\cdot\(\v{S}_j\times\dot{\v{S}}_j\) -\v{S}_j\cdot\(\v{S}_i\times\dot{\v{S}}_i\) +h.c.\]\end{align}
where we have explicitly forced $e_{ij}$ to be Hermitian in order to avoid operator ordering ambiguities in restoring the path-integral expression to operator form.  The time-evolution of $\v{S}_i$ follows from evolving under the Heisenberg Hamiltonian $H=J\sum_{\<ij\>}\v{S}_i\cdot\v{S}_j$:
\begin{align} \label{eq:TimeEvolution}\dot{\v{S}}_i = -i[H,\v{S}] = J\sum_{\v{d}}\v{S}_{i+\v{d}}\times \v{S}_i\end{align}
where $\v{d}$ are the vectors connecting site $i$ to its nearest neighbors.  Combining Eqs.~\ref{eq:TimeEvolution} and \ref{eq:SpaceTimeWilsonLoop} gives:
\begin{align}\label{eq:EmergentElectricField} e_{ij} = \frac{J}{2}\[\sum_{d_j}\v{S}_i\cdot\v{S}_{j+d_j} -\sum_{d_i}\v{S}_j\cdot\v{S}_{i+d_i}\] \end{align}
where $d_i,d_j$ are the nearest neighbor vectors of sites $i$ and $j$ respectively.

\begin{figure}[ttt]
\begin{center}
\includegraphics[width = 1.5in]{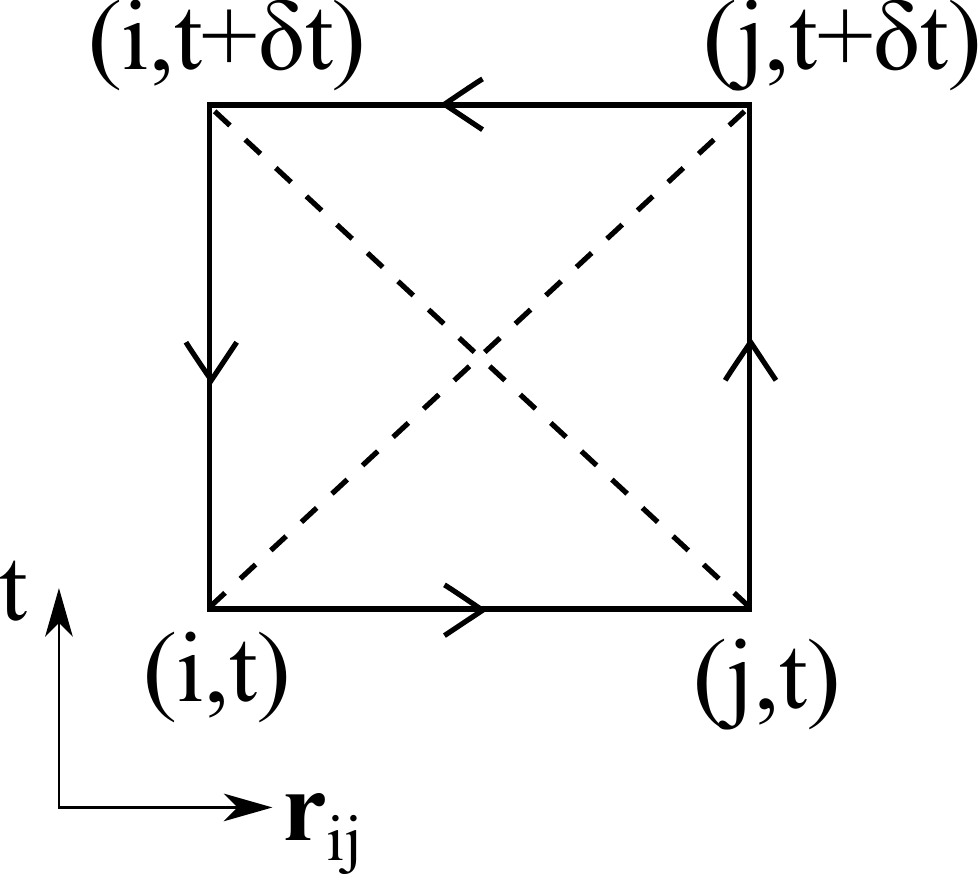}
\end{center}
\vspace{-.2in}
\caption{Space-Time Wilson loop corresponding to the emergent electric field $e_{ij}$ on lattice-bond $\<ij\>$. }
\vspace{-.2in}
\label{fig:SpaceTimePlaquette}
\end{figure}

For a 3-site triangle, Eq.~\ref{eq:EmergentElectricField} reduces to a discrete gradient of bond-energy (on the lattice of bonds), which was shown by Bulaevskii et al. to be proportional to the polarization arising from virtual charge fluctuations.

\subsection{Physical Meaning of the Emergent Gauge Field in the Presence of DM Interactions, and Field Induced Ordered States}

In the presence of DM interactions, the above derivation of the physical meaning of $b$ and $\v{e}$ is altered.  As described previously, since the in-plane DM vectors vanish within the unit cell, the uniform $z-$component of the DM interactions is expected to be the most important. On general symmetry grounds, the DM terms allow $b_{ijk}\rightarrow \v{S}_i\cdot\(\v{S}_j\times\v{S}_k\)+\alpha \(S_i^z+S_j^z+S_k^z\)$, where $\alpha$ is of order: $\alpha \approx \mathcal{O}\(\frac{D}{J}\)$.  The precise form of $\alpha$ may be derived from the Wilson loop derivation outlined above but incorporating the DM interactions at the slave-fermion mean-field level as a spin-orbit coupling term for the spinons.  

The appearance of this new term shows directly that any perturbation that couples to the z-magnetization will tend to induce an emergent-gauge-magnetic flux. For example, as discussed in further detail in Ref.~\onlinecite{LeeNagaosa}, these considerations demonstrate that neutron scattering experiments can in principle be used to detect gapless fluctuations in $b$.  

Furthermore, it has implications for the behavior of Herbertsmithite in an applied magnetic field. For example, the Zeeman term $\frac{g\mu_B}{2}H^z$ from an external field $H^z$ along z will induce a magnetization $M^z=\chi_sH^z$, which will produce a corresponding gauge magnetic flux $b_\text{ind}\sim\alpha\chi_SH^z$.  In perfectly clean Dirac spin-liquid, the density of states vanishes at the Dirac point, and the spin susceptibility grows linearly with $H$: $\chi_S^\text{Dirac}\sim \frac{H^z}{J^2}$.  Empirically, however, Herbertsmithite appears to have spin-susceptibility that tends to a constant of order a fraction of $\chi_S\sim \frac{1}{J}$ at low temperature\cite{Helton,Imai,Olariu}.  Within the U(1) spin-liquid scenario, one could possibly interpret this finite susceptibility as a finite density of states induced, for example, by impurity scattering.

In this case, the induced magnetization per spin from an external field $H^z$ would be: $M^z\sim \frac{H^z}{J}$, producing an emergent gauge flux through each triangle of order $\Phi_b = ba^2\sim\frac{DH^z}{J^2}\sim 10^{-3} H^zT^{-1}$.  This emergent gauge flux has a magnetic length $\ell_b\sim 30a\sqrt{\frac{H^z}{1\text{ Tesla}}}\sim 10^2\AA \sqrt{\frac{H^z}{1\text{ Tesla}}}$, which corresponds roughly to the ordinary orbital magnetic length an electron gas in the field $H^z$.  Therefore we expect that in a U(1) spin-liquid, the spinons will see this induce $b$ flux as an orbital field, with an effective $b$-flux density that is comparable the physical flux density of $H^z$.  This is effect is surprisingly large, since charge motion is largely frozen deep in the Mott insulating phase, and purely electronic mechanisms give effective flux for spinons that is smaller factor of $\(\frac{t}{U}\)^3\sim 10^{-3}$ compared to the physical flux\cite{Motrunich}.

This large orbital $b$ field, would then induce Landau levels into the Dirac spectrum.  Neglecting Zeeman splitting, and in the absence of interactions would have a four-fold degenerate zeroth Landau level, with an approximate SU(4) symmetry as in single-layer graphene\cite{Nomura,Alicea}.  However, the infinite density of states will generically cause interactions to remove this degeneracy and split the zeroth Landau level, gapping out the spinons and allowing instantons to proliferate.  The nature of the resulting broken symmetry state will depend on the manner in which the Landau-level symmetry is broken, as different breakings of the SU(4) symmetry result in different quantum Hall responses (e.g. quantum- spin, valley, etc... hall effects).  Within the Dirac spin-liquid scenario, such effects could possibly account for the field induced spin-freezing past a critical external field strength of a couple Tesla, which has been observed in Herbertsmithite through NMR measurements\cite{FieldInducedSpinFreezing}. Moreover, this scenario could also be relevant in a weakly gapped $Z_2$ spin-liquid scenario, where the $Z_2$ state arises out of the Dirac spin-liquid state by BCS pairing.  In such a scenario, the minimal field required to induce spin-ordering could possibly be interpreted as $H_{c}$ for the spinon superconductor (one would naively expect spinon superconductors formed by adding pairing to a $U(1)$ spin-liquid to be strongly Type-1 due to the large effective gauge-charge for the emergent gauge field).

At strong-magnetic fields where the Zeeman-energy determines splitting of the lowest Dirac Landau-level, there will be a quantum spin-Hall response\cite{Abanin}, tying a spin-flip to each instanton insertion leading to a field-induced magnetically ordered state.  At lower fields, alternative scenarios might occur\cite{Alicea}.  For example, if the zeroth Landau-level splits by valley polarization, there will be a resulting quantum valley-Hall effect. This would endow instantons with sub-lattice symmetry breaking quantum numbers; and proliferating such instantons would result in a valence-bond crystal (VBC) state.  This raises the possibility of a various sequences of continuous phase transitions as a function of field, for instance from Dirac-spin-liquid to low-field VBC state, to a high-field magnetically ordered state.

The emergent gauge electric field will also be altered by the presence of DM interactions.  On symmetry grounds, $\v{e}$ can receive contributions of order $\alpha$ from operators like $\(\v{S}_i\times\v{S_j}\)$, which correspond to spin-current in the spinon theory.  However, unlike the magnetic field case described above, it is difficult to externally couple to spin currents, so it is unclear whether these extra terms enable useful new probes.

\section{Slave Rotor Effective Field Theory \label{sec:AppSlaveRotor}}
In this paper, we have used the strong-coupling $t/U$ expansion of Bulaevskii et al.\cite{Bulaevskii} to incorporate virtual charge fluctuations into the effective spin-model description of the Hubbard model.
An alternative approach that was previously used in Refs.~\onlinecite{IoffeLarkin,NgLee}, is to work directly with the Hubbard model using a slave-boson or slave-rotor effective theory.  However, as we will show here, while qualitatively similar, the mean-field treatment of the slave-rotor theory does not capture the correct $t/U$ dependence of charge-excitations in a strong Mott insulator.  To show this fact, we begin by reviewing the slave-rotor framework and its predictions for the electromagnetic response of a U(1) spin-liquid Mott insulating state.

The slave-rotor approach begins by re-writing the physical electron operator at site $i$ and with spin $\sigma$ as a product of a spin-less boson $b_i$ and spinful fermionic spinon $f_{i\sigma}$: $c_{i,\sigma} = b_if_{i,\sigma}$, subject to the constraint: $n_b=n_f-1$.  It is technically convenient to describe the boson as a U(1) quantum rotor: $b_i = e^{i\theta_i}$. This description suffers from a U(1) gauge redundancy corresponding to the local transformations: $\theta_i\rightarrow \theta_i+\Lambda_i$, $f_{i,\sigma}\rightarrow e^{-i\Lambda_i}f_{i,\sigma}$. As a result the corresponding effective theory will contain a fluctuating compact U(1) gauge field, $a^\mu_i$.

By a choice of gauge, one can associate the physical electron charge with $b$, and a Mott insulating state is then described by a charge gap $\Delta$ for the bosonic rotor-excitations.  The resulting low-energy effective theory reads:
\begin{align} S&=\int d\tau \(L_b+L_f\)
\nonumber \\
\label{eq:SlaveRotorMFAction} L_b =& \sum_i\frac{|\(\partial_\tau+a_i^0+A^0_i\)b_i|^2+\Delta^2|b_i|^2}{2U}-
\nonumber\\
&-\sum_{\<ij\>}t^b_{ij}e^{i(a_{ij}+A_{ij})}\bar{b}_ib_j 
\nonumber \\
L_f &= \sum_i \bar{f}_i\(\partial_\tau-\mu-a^0_i\)f_i-\sum_{\<ij\>}t^f_{ij}e^{ia_{ij}}\bar{f}_if_j\end{align}
where $A^\mu_i$ is the physical electromagnetic field, $\mu$ is determined such that $\<n_b\>=\<n_f\>-1$, and where we have treated $b_i$ as a soft-rotor by relaxing the rigid constraint $|b_i|^2=1$.

The hopping amplitudes $t^f$ and $t^b$ can be estimated within a mean-field approximation: $t^b_{ij} \approx t_{ij}\<\bar{f}_if_j\>$ and $t^f_{ij}\approx\<e^{i(\theta_i-\theta_j)}\>$.  Whereas $t^b\approx t$ independent of interaction strength $U$, the spinon bandwidth decreases stronger interactions starting from $t^f\approx t$ in the vicinity of the Mott transition ($\Delta\ll t,U$), and reducing to $t^f\rightarrow J \approx \frac{t^2}{U}$ for a strong Mott insulator ($U\gg t$).  In the large $U$ limit, we see that the slave-rotor effective action simply reproduces that of the Schwinger fermion treatment of the Heisenberg spin-model, without discarding the gapped charge fluctuations that enable non-zero optical conductivity.

Here we see the first hint that the mean-field treatment of the slave-rotor theory does not quantitatively capture the dynamics of charge excitations: the mean-field treatment predicts that charge excitations propagate with effective mass $\frac{1}{2ta^2}$, whereas, in a strong Mott insulator one would expect strong coupling between spin fluctuations to enhance the holon mass to $\sim\(\frac{1}{Ja^2}\)$.\cite{KaneLeeRead}

\subsubsection{Speed of Emergent Light}
The Maxwell term for $a$ generated by integrating out the gapped boson fields in the slave-rotor action of Eq.~\ref{eq:SlaveRotorMFAction}, takes the form $\epsilon \v{e}^2+\frac{1}{\mu} b^2$, where $\v{e} = -\nabla a_0+i\partial_\tau \v{a}$ and $b = \hat{z}\cdot\nabla\times\v{a}$ are the emergent electric and magnetic fields respectively, and $\epsilon$ and $\mu$ are the corresponding dielectric constant and magnetic permeability.  

%The speed of light, $c^*$ for the emergent gauge field can be identified by comparing two gauge configurations corresponding to the same constant electric field, $\v{e}_0$: 1) $a_0(\v{r},\tau) = -\frac{e_0}{iq}e^{iq\hat{e}_0\cdot r}$, $\v{a}=0$ and 2) $a_0 = 0$, $\v{a}= -\frac{c^*\v{e}_0}{-i\omega}e^{-i\omega\tau}$, where $q,\omega\ll \Delta,t,U$.  The corresponding action-densities are: $\frac{K_b^{00}(\omega=0,q)}{q^2}\v{e}_0^2$ and $\frac{(c^*)^2K_b^{ii}(\omega,q=0)}{\omega^2}\v{e}_0^2$.  By gauge invariance these expressions are equal to each-other and to $\epsilon\v{e}_0^2$, which fixes:
%\begin{align} \label{eq:SpeedofLight} c^* = \sqrt{\frac{\omega^2}{q^2}\frac{K_b^{00}(\omega=0,q\rightarrow 0)}{K_b^{ii}(\omega\rightarrow 0,q=0)}} \end{align}

We can compute the speed of emergent `light', $c^*=\frac{1}{\sqrt{\epsilon\mu}}$, within the slave-rotor theory for a strong Mott insulator.  The dielectric constant $\epsilon$ can be determined by examining the action cost of a uniform electric field corresponding to $a_0=0$, $\v{a} = \frac{\v{e}_0}{\Omega}e^{-i\Omega\tau}$:
\begin{align} K_b^{ii}&(\Omega,q=0) = 
\nonumber\\
&=\sum_{\omega,k}\frac{\partial_{k^i}\e_k}{\frac{(\omega+\Omega)^2+\Delta^2}{2U}+\e_{k}}\frac{\partial_{k^i}\e_k}{\frac{\omega^2+\Delta^2}{2U}+\e_{k}}-K_b^{ii}(0,0)
\nonumber\\
&\approx -\frac{3\Omega^2}{4U^3}\sum_k\(\frac{\partial \e_k}{\partial k^i}\)^2 \sim \frac{t^2a^2}{U^3}\Omega^2\end{align}
%\begin{align} K_b^{00}&(\omega=0,q) = K_b^{00}(\Omega=0,q)-K_b^{00}(0,0) 
%\nonumber\\
%&= \sum_{\omega,k}\frac{1}{\frac{\omega^2+\Delta^2}{2U}+\e_{k+q}}\frac{1}{\frac{\omega^2+\Delta^2}{2U}+\e_{k}}-K_b^{00}(0,0)
%\nonumber\\
%&\approx-(2U)^3\sum_{\omega,k} \frac{\frac{1}{2}q^iq^j\partial_{k^i}\partial_{k^j}\e_{k}}{\(\omega^2+\Delta^2\)^3}
%\nonumber\\
%&\approx\frac{3}{2U^2}\sum_{k} \frac{1}{2}q^iq^j\(-\frac{\partial^2\e_{k}}{\partial{k^i}\partial{k^j}}\)
%\sim \frac{ta^2}{U^2}q^2
%\end{align}
where we have kept only the leading order contributions in $t/U$ and identified $\Delta\approx U$, as appropriate for a strong-Mott insulator. In the first line, we have used the fact that the diamagnetic contribution to $K_b$ is equivalent to explicitly subtracting $K_b^{00}(0,0)$; this just ensures that $K_b(0,0)=0$ as required by gauge invariance.  This result identifies: $\epsilon\sim \frac{t^2a^2}{U^3}$.

%Inserting these results into Eq.~\ref{eq:SpeedofLight} gives:
%\begin{align} c^* \approx \sqrt{U\frac{\sum_k\hat{q}^i\hat{q}^j\(-\partial_{k^i}\partial_{k^j}\e_k\)}{\sum_k \(\partial_{k^i}\e_k\)^2}}\sim \sqrt{\frac{U}{t}} \end{align}

The magnetic permeability, $\mu$ is obtained from $K_b^{ij}(\Omega=0,q\rightarrow 0)$, which by gauge invariance is proportional to $q^2-q^iq^j$.  Therefore, it is sufficient to compute only the $K_b^{ii}$ component:
\begin{align} K_b^{ii}&(\Omega=0,q)
\nonumber\\
&= \sum_{\omega,k}\frac{\(\frac{\partial\e_k}{\partial k^i}\)^2} {\[\frac{\omega^2+\Delta^2}{2U}+\e_{k+q/2}\]\[\frac{\omega^2+\Delta^2}{2U}+\e_{k-q/2}\]}
-K_b^{ii}(0,0)\end{align}
Expanding the denominator to $\mathcal{O}(q^2)$ gives:
\begin{align} K_b^{ii}&(\Omega=0,q)
\nonumber\\
&\approx -q^lq^m\sum_{\omega,k}\frac{\(\frac{\partial\e_k}{\partial k^i}\)^2} {\(\frac{\omega^2+\Delta^2}{2U}\)^2}\[\frac{\frac{1}{4}\frac{\partial\e_{k}}{\partial k^l\partial k^m}}{\(\frac{\omega^2+\Delta^2}{2U}\)}+\frac{\frac{3}{4}\frac{\partial\e_k}{\partial k^i}\frac{\partial\e_l}{\partial k^m}}{\(\frac{\omega^2+\Delta^2}{2U}\)^2}\]
\nonumber\\
&\approx -12U^4q^lq^m\[\sum_k \(v^i\)^2v^lv^m\]\[\sum_\omega\frac{1}{\(\omega^2+U^2\)^4}\]
\end{align}
where the first term in the second line is the average of the band-mass over the full band, which vanishes, and $v_i\equiv \frac{\partial \e_k}{\partial k^i}$ is the velocity.  For (nearly) isotropic bands, we can approximate the velocity average by $\sim v_F^4\delta_{lm}$, and we find that 
\begin{align} K_b^{ii}&(\Omega=0,q\rightarrow 0)\sim \frac{(ta)^4}{U^3} \end{align}
This gives: $\epsilon\sim \frac{t^2a^2}{U^3}$, and $\frac{1}{\mu}\sim \frac{(ta)^4}{U^3}$.  Combining these results gives: \begin{align}c^*=\frac{1}{\sqrt{\epsilon\mu}}\sim ta \sim v_F\end{align}
where $v_F$ is the bare Fermi velocity corresponding to half-filling and hopping $t$.  Again, in a strong Mott insulator, one would expect the only relevant scale to be the spin-exchange $J$, not $t$, further fueling our suspicions that the slave-rotor mean-field is not sufficient for a quantitative description of charge dynamics.

\subsubsection{Physical Electric Current Operator}
The physical electrical current is defined as the variation of the effective lagrangian density with respect to the electromagnetic vector potential:  $j_\text{ph} = \frac{\partial\mathcal{L}}{\partial A^\mu}$.  Following the previous section, the effective action for the emergent gauge field in the presence of an external physical electrical field, after integrating out the gapped boson fields takes the form: $\mathcal{L}_E=\epsilon\(\v{e}+\v{E}\)^2$, which contains a linear coupling of the emergent and physical electric fields: $\mathcal{L}_{Ee} = 2\epsilon\v{e}\cdot\v{E}$, which, in the gauge where $\v{E} = -\partial_\tau\v{e}$ is equivalent to (after integrating by parts): $\mathcal{L}_{Ee} = -2\epsilon\partial_\tau\v{e}\cdot\v{A}$.  Meaning that the physical electrical current is:
\begin{align} \v{j}_\text{ph} = 2e\epsilon\partial_\tau\v{e} \sim e\frac{t^2a^2}{U^3}\partial_\tau\v{e} \end{align}
Comparing to Eq.~\ref{eq:PhysicalEMCurrent}, we see that the slave-rotor theory is a factor of $t/U$ different than the strong-coupling expansion, which we attribute to the aforementioned shortcomings of the mean-field slave-rotor treatment of the gapped charge dynamics.

%\section{$t/U$ Expansion for 3-Site Hubbard Model}


\begin{thebibliography}{99}
\bibitem{Elser1}
C. Zeng and V. Elser, 
\textit{Numerical studies of antiferromagnetism on a Kagome net.}
Phys. Rev. B {\bf 42}, 8436 (1990).

\bibitem{Marston}
J.B. Marston and C. Zeng, 
\textit{Spin-Peierls and spin-liquid phases of Kagome quantum antiferromagnets.}
J. Appl. Phys. {\bf 69}, 5962 (B) (1991).

\bibitem{SachdevSPN}
S. Sachdev, 
\textit{Kagome- and triangular-lattice Heisenberg antiferromagnets: Ordering from quantum fluctuations and quantum-disordered ground states with unconfined bosonic spinons}
Phys. Rev. B {\bf 45}, 12377 (1992).

\bibitem{Elser2}
P.W. Leung and V. Elser, 
\textit{Numerical studies of a 36-site kagome antiferromagnet.}
Phys. Rev. B {\bf 47}, 5459 (1993)

\bibitem{Lecheminant}
P. Lecheminant, B. Bernu, C. Lhuillier, L. Pierre, and P. Sindzingre, 
\textit{Order versus disorder in the quantum Heisenberg antiferromagnet on the kagome lattice using exact spectra analysis.}
Phys. Rev. B {\bf 56}, 2521 (1997).

\bibitem{YanHuseWhite}
S. Yan, D. A. Huse, and S. R. White, 
\textit{Spin-Liquid Ground State of the S = 1/2 Kagome Heisenberg Antiferromagnet.}
Science {\bf 332}, 1173 (2011).

\bibitem{Shores}
M.P. Shores, E.A. Nytko, B.M. Barlett, and D.G. Nocera,
\textit{A structurally perfect S= 1/2 kagome antiferromagnet.}
J. Am. Chem. Soc. 127, 13462 (2005).

\bibitem{Helton}
J.S. Helton, et al.
\textit{Spin Dynamics of the Spin-1/2 Kagome Lattice Antiferromagnet ZnCu$_3$(OH)$_6$Cl$_2$.}
Phys. Rev. Lett. {\bf 98}, 107204 (2007).

\bibitem{Imai}
T. Imai, E. A. Nytko, B.M. Bartlett, M.P. Shores, and D. G. Nocera
\textit{$^{63}$Cu, $^{35}$Cl, and $^1$H NMR in the S=1/2 Kagome Lattice ZnCu$_3$(OH)$_6$Cl$_2$.}
Phys. Rev. Lett. {\bf 100}, 077203 (2008).

\bibitem{Olariu}
A. Olariu, P. Mendels, F. Bert, F. Duc, J.C. Trombe, M.A. de Vries, and A. Harrison,
\textit{$^{17}$O NMR Study of the Intrinsic Magnetic Susceptibility and Spin Dynamics of the Quantum Kagome Antiferromagnet ZnCu$_3$(OH)$_6$Cl$_2$.}
Phys. Rev. Lett. {\bf 100}, 087202 (2008).

\bibitem{NeutronContinuum}
Tianheng Han et al., to be published.

\bibitem{RanVMC}
Y. Ran, M. Hermele, P. A. Lee, and X.-G.Wen, 
\textit{Projected-Wave-Function Study of the Spin-1/2 Heisenberg Model on the Kagome Lattice.}
Phys. Rev. Lett. {\bf 98}, 117205 (2007).

\bibitem{Becca}
Y. Iqbal, F. Becca, S. Sorella, and D. Poilblanc,
\textit{Gapless spin-liquid phase in the kagome spin-1/2 Heisenberg antiferromagnet.}
arXiv:1209.1858 (2012).

\bibitem{HermeleProperties}
M. Hermele et al. 
\textit{Properties of an Algebraic spin-liquid on the Kagome lattice}
Phys. Rev. B {\bf 77}, 224413 (2008)

\bibitem{WhiteJ1J2}
White, S. R. 
\textit{The spin liquid ground state of the S=1/2 Heisenberg model
on the kagome lattice.}
Bull. Am. Phys. Soc. 57 MAR.L19.1 (2012); available at
http://meetings.aps.org/link/BAPS.2012.MAR.L19.1.

\bibitem{IoffeLarkin}
L.B. Ioffe, A.I. Larkin
\textit{Gapless Fermions and Gauge Fields in Dielectrics.}
Phys. Rev. B (1989)

\bibitem{LeeNagaosaNeutron}
P.A. Lee, N. Nagaosa
\textit{A Proposal to Use Neutron Scattering to Measure Scalar Spin Chirality Fluctuations in Kagome Lattices.}
arXiv:1210.3051 (2012)

\bibitem{Elsasser}
S. Elsasser, D. Wu, M. Dressel, J.A. Schlueter,
\textit{Power-law dependence of the optical conductivity observed in the quantum spin-liquid compound $\kappa$-(BEDT-TTF)2$_2$Cu$_2$(CN)$_3$.}
Phys. Rev. B {\bf 86}, 155150 (2012).

\bibitem{NgLee}
T.K. Ng, P.A. Lee, 
\textit{Power-Law Conductivity inside the Mott Gap: Application to $\kappa$-(BEDT-TTF)$_{2}$Cu$_{2}$(CN)$_{3}$}
Phys. Rev. Lett. (2007)

\bibitem{Pilon}
D.V. Pilon, C.H. Lui, T. Han, D.B. Shrekenhamer, A.J. Frenzel, W.J. Padilla, Y.S. Lee, and N. Gedik
\textit{Spin Induced Optical Conductivity in the Spin Liquid Candidate Herbertsmithite}
arXiv:1301.3501 (2013)

\bibitem{Bulaevskii}
L.N. Bulaevskii, C.D. Batista, M.V. Mostovoy, and D.I. Khomskii
\textit{Electronic orbital currents and polarization in Mott insulators}
Phys. Rev. B (2008)

\bibitem{SachdevDM}
Y. Huh, L. Fritz, and S. Sachdev
\textit{Quantum criticality of the kagome antiferromagnet with Dzyaloshinskii-Moriya interactions}
Phys. Rev. B {\bf 81}, 144432 (2010) 

\bibitem{Chubukov}
A. Chubukov, S. Sachdev, and T. Senthil, 
\textit{Quantum phase transitions in frustrated quantum antiferromagnets}
Nucl. Phys. B {\bf 426}, 601 (1994).

\bibitem{Ran}
Y.-M. Lu and Y. Ran
\textit{$Z_2$ spin liquid and chiral antiferromagnetic phase in the Hubbard model on a honeycomb lattice}
Phys. Rev. B, {\bf 84}, 024420 (2011)

\bibitem{LudwigQHTransition}
A.W.W. Ludwig, M.P.A. Fisher, R. Shankar, and G. Grinstein
\textit{Integer quantum Hall transition: An alternative approach and exact results}.
Phys. Rev. B, {\bf 50}, 7526 (1994)

\bibitem{Lurie}
see e.g. N. A. Lurie, D. L. Huber, and M. Blume, 
\textit{Computer studies of spin and energy transport in one-dimensional Heisenberg magnets.}
Phys. Rev. B {\bf 9}, 2171 (1974)
and references therein

\bibitem{SachdevBook}
S. Sachdev, 
\textit{Quantum Phase Transitions}
(Cambridge University Press, Cambridge, 1999).

\bibitem{SachdevHolography}
R. C. Myers, S. Sachdev, and A. Singh, 
\textit{Holographic Quantum Critical Transport without
Self-Duality} 
Phys. Rev. D {\bf 83}, 066017 (2011)

\bibitem{LeeNagaosa}
P.A. Lee and N. Nagaosa
\textit{Gauge theory of the normal state of high-$T_c$ superconductors}
Phys. Rev. B, {\bf 46}, 5621 (1992)

\bibitem{WenWilczekZee}
X.G. Wen, F. Wilczek, and A. Zee
\textit{Chiral spin states and superconductivity}
Phys. Rev. B, {\bf 39}, 11413 (1989)

\bibitem{OperatorOrderingEndNote}
To restore this equal-time product of Grassman fields to operator form, in principle, one needs to resolve an operator ordering ambiguity.  However, for determining the flux through a triangular loop the choice of orderings turns out to be unimportant.  For larger loops however, the ordering convention makes some difference.  In particular, the choice of Refs.~\onlinecite{LeeNagaosa,WenWilczekZee} leads to an expression for the U(1) Wilson loop that depends explicitly on the choice of base-point.  This undesireable feature can be avoided by averaging over choice of base-points.

\bibitem{Motrunich}
O. Motrunich
\textit{Orbital magnetic field effects in spin liquid with spinon Fermi sea: Possible application to $\kappa-(ET)_2Cu_2(CN)_3$.}
Phys. Rev. B {\bf 73}, 155115 (2006)

\bibitem{Nomura}
K. Nomura and A.H. MacDonald
\textit{Quantum Hall Ferromagnetism in Graphene}
Phys. Rev. Lett. {\bf 96}, 256602 (2006).

\bibitem{Alicea}
J. Alicea and M.P.A. Fisher,
\textit{Graphene integer quantum Hall effect in the ferromagnetic and paramagnetic regimes.}
Phys. Rev. B {\bf 74}, 075422 (2006)

\bibitem{FieldInducedSpinFreezing}
M. Jeong, F. Bert, P. Mendels, F. Duc, J. C. Trombe, M. A. de Vries, and A. Harrison,
\textit{Field-Induced Freezing of a Quantum Spin Liquid on the Kagome Lattice}.
Phys. Rev. Lett. {\bf 107}, 237201 (2011)

\bibitem{Abanin}
D.A. Abanin, P.A. Lee, and L.S. Levitov, 
\textit{Spin-Filtered Edge States and Quantum Hall Effect in Graphene.}
Phys. Rev. Lett. {\bf 96}, 176803 (2006) 

\bibitem{KaneLeeRead}
C.L. Kane, P.A. Lee, and N. Read
\textit{Motion of a single hole in a quantum antiferromagnet}
Phys. Rev. B. {\bf 39} 6880 (1989)

\end{thebibliography}
\end{document}